\documentclass[reprint,aps,prl,superscriptaddress,noeprint]{revtex4-2}

\usepackage[normalem]{ulem}
\usepackage{amsmath}
\usepackage{amsfonts}
\usepackage{amssymb} 
\usepackage{graphicx}
\usepackage{float}
\usepackage[caption=false]{subfig}
\usepackage{hyperref}
\hypersetup{
	colorlinks=true,       
	linkcolor=blue,          
	citecolor=green,       
	filecolor=magenta,      
	urlcolor=blue          
}
\usepackage{siunitx}
\usepackage{booktabs}
\usepackage{multirow}

\usepackage[dvipsnames]{xcolor}

\begin{document}
\title{Data-insensitive cooling of polar molecules with Rydberg atoms}	

\author{Jeremy T. Young}
\email[Corresponding author: ]{j.t.young@uva.nl}
\affiliation{Institute of Physics, University of Amsterdam, 1098 XH Amsterdam, the Netherlands}

\author{Ron Belyansky}
\affiliation{Pritzker School of Molecular Engineering, University of Chicago, Chicago, IL 60637, USA}

\author{Kang-Kuen Ni}
\affiliation{Department of Chemistry and Chemical Biology, Harvard University, Cambridge, Massachusetts 02138, USA}
\affiliation{Department of Physics, Harvard University, Cambridge, Massachusetts 02138, USA}
\affiliation{Harvard-MIT Center for Ultracold Atoms, Cambridge, Massachusetts 02138, USA}

\author{Alexey V. Gorshkov}
\affiliation{Joint Quantum Institute and Joint Center for Quantum Information and Computer Science, NIST/University of Maryland, College Park, Maryland 20742, USA}

\date{\today}

\begin{abstract}

We propose a method to sympathetically cool polar molecules with Rydberg atoms without destroying the quantum information encoded in the polar molecules. While the interactions between the two are usually state-dependent, we show how to engineer state-insensitive interactions between the hot molecules and the cold atoms with a suitable choice of internal states and the application of external fields. The resulting interactions, which may be van der Waals or dipolar, induce a phonon swap interaction between the two species, thereby coherently cooling the polar molecules without affecting the internal state, a process which can be repeated if the atoms are cooled again or new cold atoms are brought in. Our cooling schemes open the possibility of extending quantum computation and simulation times in emerging hybrid tweezer arrays of polar molecules and neutral atoms.

\end{abstract}


\maketitle

In the past decade, rapid progress has been achieved in the utilization of both polar molecules and neutral atoms for applications in quantum technologies. In particular, arrays of individually trapped neutral atoms have shown to be a versatile and highly controllable platform for quantum simulation and quantum computation \cite{Endres2016,Barredo2016,Browaeys2020,Kaufman2021,Bluvstein2024,Manetsch2024}. More recently, significant experimental advances have been made in tweezer arrays of polar molecules \cite{Liu2018,He2020a,Zhang2022,Anderegg2019,Burchesky2021,Bao2023,Holland2023b,
Ruttley2023,Guttridge2023,Park2023,Ruttley2024,Picard2024,Lu2024,Bao2024,Picard2024a, Ruttley2025}, opening the way to polar-molecule-based quantum technologies \cite{Cornish2024}. However, whether one is considering polar molecules or neutral atoms, the effects of heating of motional degrees of freedom place a crucial limitation on the experimental timescales that can be realized. 

In recent work, it was proposed that Rydberg interactions can be used to cool the motional degrees of freedom in neutral atoms \cite{Belyansky2019}---such as can occur due to shaking of the atomic array from laser intensity noise \cite{Savard1997}, mechanical forces from Rydberg interactions \cite{Jaksch2000,Saffman2005,Saffman2010}, or incoherent light scattering \cite{Pichler2010}---without destroying quantum information stored in the internal states. Inspired by sympathetic cooling of trapped ions where interactions only depend on the charge and not the internal state \cite{Barrett2003,Reichenbach2007}, Ref.~\cite{Belyansky2019} demonstrated how analogous state-insensitive van der Waals (vdW) interactions can be realized with Rydberg atoms. Thus by preparing a set of cold auxiliary atoms, hot atoms can be sympathetically cooled by the auxiliary atoms, which can then be cooled again to realize further sympathetic cooling. 

In this work, we extend these ideas to hybrid tweezer arrays of neutral atoms and polar molecules, which have been the subject of recent theoretical and experimental interest \cite{Zhang2022a,Wang2022,Guttridge2023}. We use the ability to tune the Rydberg transition (via state choice or external fields) into near-resonance with the molecular transition and enhance molecule-Rydberg dipolar interactions, a feature which has been similarly used in proposals for quantum computation \cite{Kuznetsova2011,Zhang2022a,Wang2022}, (state-sensitive) sympathetic cooling \cite{Zhao2012} and laser cooling \cite{Huber2012}, state detection \cite{Kuznetsova2016,Zeppenfeld2017,Patsch2022}, and mediating interactions \cite{Kuznetsova2018}, as well as experimental observations of dipolar exchange \cite{Zhelyazkova2017b,Jarisch2018,Gawlas2020,Zhu2025}. Although not the focus here, charge-dipole interactions at short scales have similarly been considered for blockade \cite{Guttridge2023} and (state-sensitive) sympathetic cooling \cite{Zhang2024a}.

\begin{figure}
\centering
\includegraphics[scale=.5]{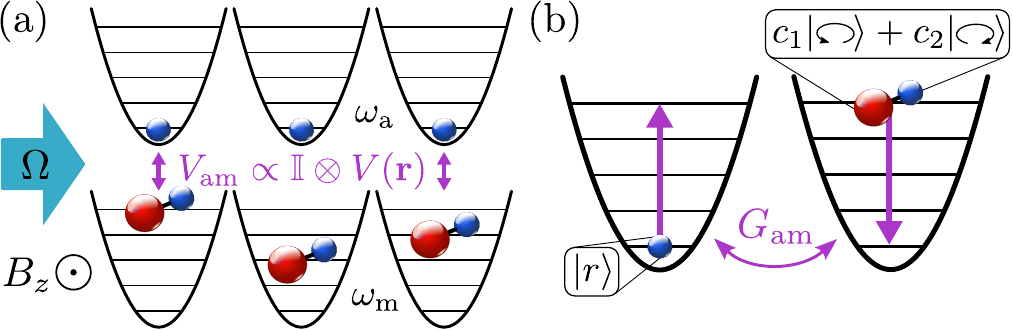}
\caption{Schematic of cooling protocol. (a) Each hot polar molecule is paired with a cold atom. Through the application of external fields, the atom-molecule interactions $V_{\text{am}}$ are independent of the internal molecular states. (b) The state-insensitive interactions realize an effective phonon swap at rate $G_{\text{am}} \propto \partial^2_\mathbf{r} V_{\text{am}}$, thereby coherently cooling the molecules without affecting the molecular qubit, which is formed via internal rotational or nuclear spin states. \label{fig:schematic}}
\end{figure}

We develop two primary approaches to sympathetic data-insensitive cooling of polar molecules with Rydberg atoms 
via coherent phonon exchange (see Fig.~\ref{fig:schematic}). We utilize the enhanced molecule-Rydberg interactions from the strong Rydberg transition dipole moments, the ability to easily cool neutral atoms, and the ability to realize state-insensitive molecule-Rydberg interactions, providing a new avenue for preparing polar molecules in the motional ground state and extending quantum simulation and computation timescales. In the first approach (vdW cooling), we show how naturally-emerging vdW interactions can achieve this. In the second approach (dipolar cooling), we utilize strong microwave dressing to engineer the necessary interactions via strong dipolar interactions.
In both approaches, we focus on encoding the qubits in molecular rotational states in a regime where nuclear fine structure can largely be ignored. In a generalized approach (hyperfine cooling), we show how the nuclear hyperfine structure of single rotational states can be used instead for the first two approaches and, in combination with the application of a strong magnetic field, dramatically improve both the performance and the versatility of these two approaches. 
As examples, we consider cooling the bialkalis $^{23}$Na$^{133}$Cs and $^6$Li$^{133}$Cs with either of their constituent atoms.

\textit{Phonon exchange}.---The general mechanism for the cooling is a combination of state-insensitive interactions and phonon exchange. By engineering state-insensitive interactions of the form $\hat{V} = \mathbb{I}_{\text{internal}}
\otimes V(\hat{\mathbf{r}})$, we can neglect the internal states of the molecule and atom and express our Hamiltonian purely in terms of the phonons. Consider a single atom/molecule pair's quadratic Hamiltonian for fluctuations in the $z$-direction
\begin{subequations}
\begin{multline}
\hspace{-.35cm} \hat H_{z,2} = \omega_{\text{a},z} \hat a_z^\dagger \hat a_z + \omega_{\text{m},z} \hat m_z^\dagger \hat m_z - G_{\text{am},z} (\hat a_z + \hat a_z^\dagger)(\hat m_z + \hat m_z^\dagger) + \\ G_{\text{a},z} (\hat a_z + 
\hat a_z^\dagger)^2 + G_{\text{m},z} (\hat m_z + \hat m_z^\dagger)^2,
\end{multline}
\begin{equation}
G_{\text{am}} = \frac{\partial_z^2 V(\mathbf{r}_{\text{am}})}{\sqrt{4 M_a M_m \omega_a \omega_m}}, \quad
G_{\text{a/m}} = \frac{\partial_z^2 V(\mathbf{r}_{\text{am}})}{4 M_{\text{a/m}} \omega_{\text{a/m}}},
\end{equation}
\end{subequations}
where $\hat a_z$ ($\hat m_z$) denotes the atomic (molecular) phonon annihilation operator for the $z$-direction, $\omega_{\text{a/m},z}$ are the trapping frequencies in the $z$-direction, and $M_{\text{a/m}}$ the masses, where a/m subscripts denote the atom or molecule. We have assumed that $z$ is orthogonal to the particle separation $\mathbf{r}_{\text{am}}$, and higher-order terms are dropped under the assumption $(M \omega \mathbf{r}_{\text{am}}^2)^{-1} \ll 1$, which controls the perturbative expansion \cite{suppcite}. This ensures that there are no linear terms, although the method we describe here can be generalized for the presence of linear terms as well---and hence to the case where $z$ is along  $\mathbf{r}_{\text{am}}$---by turning the interactions on adiabatically \cite{Belyansky2019}. We take the $z$-axis to be our quantization axis. 

In the limit $\omega_z \gg G$, only terms which conserve total phonon number are relevant under the rotating-wave approximation. In the rotating frame of the molecular phonons, 
\begin{equation}
\hat H_{z,2} \approx \Delta_{\text{am}} \hat a^\dagger \hat a - G_{\text{am}}(\hat a^\dagger \hat m + \hat m^\dagger \hat a),
\end{equation}
where $\Delta_{\text{am}} = \omega_{\text{a},z} +G_{\text{a},z} - \omega_{\text{m},z} - G_{\text{m},z}$.
Thus, the interaction results in phonon exchange between the atoms and molecules. When $\Delta_{\text{am}} = 0$, the exchange is resonant and all phonons are swapped after time $t_{\text{swap}} = \pi/2G_{\text{am}}$. Otherwise, assuming no initial atomic phonons, the molecular phonons can be reduced by a factor $\Delta_{\text{am}}^2/(4 G_{\text{am}}^2 + \Delta_{\text{am}}^2)$ after time $t = \pi/\sqrt{4 G_{\text{am}}^2 + \Delta_{\text{am}}^2}$. Since $\omega_z \gg G$, it is necessary to ensure that $\omega_{\text{a},z} \approx \omega_{\text{m},z}$ to realize significant phonon swap; the remaining contributions from $G_{\text{a/m},z}$ can be removed by adequate modification of $\omega_{\text{a/m},z}$. Throughout the text we assume $\Delta_{\text{am}} = 0$. In general, we are interested in cooling an ensemble of molecules with an ensemble of atoms simultaneously. Although the additional interactions reduce the efficiency of the phonon exchange, the reduction is minimal and the two-particle picture holds qualitatively \cite{Belyansky2019,suppcite}. 

\emph{Molecule-Rydberg interactions}.---The rotational energies of the molecule are defined by the Hamiltonian $\hat H_{\text{rot}} = B_0 \hat{\mathbf{N}}^2$, where $\hat{\mathbf{N}}$ is the rotation operator of the molecule and $B_0$ the rotational constant. By changing the principal quantum number $n$ of the Rydberg atoms, the energy separation between nearby (in energy) Rydberg states scales as $n^{-3}$. With a careful selection of Rydberg states,  the Rydberg transition frequency can be tuned so it is similar to the molecular transition frequency, with smaller $B_0$ requiring larger $n$.
This results in enhanced vdW interactions and the ability to strongly dress both Rydberg and molecular states simultaneously for direct dipolar interactions. When utilizing hyperfine structure, it is possible to tune the transition frequencies via an external magnetic field, providing flexibility in $n$.

For concreteness, we consider two alkali dimers with relatively large dipole moments $d$: $^{23}$Na$^{133}$Cs (4.53 D) and $^6$Li$^{133}$Cs (5.36 D) \cite{Fedorov2014}. $^{23}$Na$^{133}$Cs has a relatively small $B_0 = 2 \pi \times 1.740$ GHz, while $^6$Li$^{133}$Cs has a relatively large $B_0 = 2 \pi \times 6.520$ GHz \cite{Aldegunde2017}, so a comparison of the two will illustrate the important role of $B_0$. Moreover, ground state $^{23}$Na$^{133}$Cs molecules have been successfully trapped in optical tweezers \cite{Zhang2022}. 
We focus on cooling each molecule with either of its constituent atoms. In the supplement, we show how our schemes can be extended to more complex molecules with the example of $^{40}$Ca$^{19}$F \cite{suppcite}.
For convenience, we drop the isotope superscripts in the remainder of the text.

\begin{figure}
\includegraphics[scale=1]{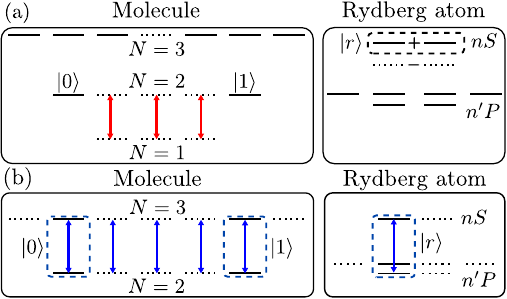}
\caption{Level diagrams used for data-insensitive cooling of polar molecules with Rydberg atoms via (a) vdW interactions and (b) microwave-dressed dipole-dipole interactions.
The arrows indicate transitions which are driven with a microwave drive. The solid (dotted) levels correspond to states which are (not) involved in the vdW or dipolar interactions, both directly and virtually. \label{fig:scheme}}
\end{figure}

\textit{vdW Cooling}.---First, we discuss how to cool the molecules using vdW interactions,  illustrated in Fig.~\ref{fig:scheme}(a). We focus on the case of an auxiliary Rydberg S state, although our results can be generalized to $L \neq 0$ states. By choosing a pair of molecular states with equal but opposite values of $m_N$, where $m_N$ is the eigenvalue of the projection $\hat N_z$ on the quantization axis, we ensure that they interact in the same way with the Rydberg state $|r\rangle$ by symmetry. Additionally, by choosing states such that $\delta m_N > 2$, we ensure that there will be no vdW interactions which exchange one molecular state for the other, since these can change $m_N$ of either state by at most $2$. The particular molecular states we use to encode the qubit
are $|0\rangle = |N = 2, m_N = -2 \rangle$ and $|1\rangle = |N = 2, m_N = 2 \rangle$. Although these states ensure no  exchange of $|0\rangle, |1\rangle$, it is possible to leave this basis, e.g.,~to $|N=2, m_N = 1\rangle$. To circumvent this, we employ a strong microwave drive to bring these transitions out of resonance. By choosing the polarization so that it only drives $\pi$ transitions with the $N=1$ states, neither qubit state is affected. 

In addition to $\pm m_N$ symmetry in the molecules, we require $m_J$ symmetry in the Rydberg state $|r\rangle$. 
The resulting molecular-Rydberg vdW interactions take the form
\begin{subequations}
\begin{equation}
V(r, \theta,\phi) = \frac{C_6}{r^6} F(\theta,\phi),
\end{equation}
\end{subequations}
where $C_6$ is the vdW dispersion coefficient which encodes the strength of the interactions and $F(\theta,\phi)$ captures the angular dependence of the Rydberg interactions. 
The form of $F(\theta,\phi)$ depends on the choice of states used in the cooling and the relevance of hyperfine structure \cite{suppcite}. For alkali atoms in an S state, $J=1/2$, we utilize the symmetric superposition $|r\rangle = (|{m_J = -1/2} \rangle + |{m_J = +1/2 }\rangle)/\sqrt{2}$. 

\textit{Dipolar cooling}.---We also consider leveraging the enhanced interaction strengths of dipolar interactions. Although the dipolar interactions between polar molecules and atoms are only near-resonant, we circumvent this via strong microwave dressing. By considering a limit where the interaction $V$ satisfies $V \ll \Omega_\text{max}$, where $\Omega_\text{max}$ is the strongest microwave drive, we can work in the dressed basis. The scale of $\Omega_\text{max}$ in this condition is set by the Rydberg Rabi frequency since it is much larger than the molecular Rabi frequency. Since the molecules and atoms are dressed with the same microwave field, one of the two transitions will be off-resonant with the drive, weakening the dressed interactions. However, by dressing the molecules resonantly, the strong Rydberg dipole moment ensures that the detuning on the Rydberg transition is not large compared to its Rabi frequency.

The scheme is illustrated in Fig.~\ref{fig:scheme}(b). Like vdW cooling, we use selection rules to prevent resonant flip-flops by setting $\delta m_N > 2$. The $|N=2 ,m_N = \pm 2\rangle $ states are resonantly dressed with the $|N=3, m_N = \pm 2 \rangle $ states. Once more, the symmetry in $m_N$ ensures state-insensitive interactions, while $m_J$ symmetry is no longer necessary provided resonant transitions to other $m_J$ are forbidden, e.g., via light shifts. 

Unlike in the case of vdW cooling, we do not limit ourselves to $\text{S}\to\text{P}$ transitions. The reason for this is twofold. First, while vdW interactions require the pair of virtual transitions to have similar but opposite transition energies, for microwave dressing the transition frequencies only need to be similar in magnitude. Second, the transition frequencies between S and P states tend to be highly ``symmetric''. By this, we mean that the transition frequencies from an S state to the nearest P states above and below in energy are similar in magnitude. In contrast, the $\text{P}\to\text{D}$ transitions are much more asymmetric, so the near-resonance condition for a transition up in energy and the condition for a transition down in energy occur at different $n$, providing increased tunability and a decreased likelihood for the drive to couple to many Rydberg states. For Li, we consider $\text{S}\to\text{P}$ transitions, which are already highly asymmetric. 

Unlike before, an additional field which couples to $N=1$ states is no longer necessary because the differing Clebsch-Gordan coefficients, and thus Rabi frequencies, for the $|m_N| \neq 2$ transitions ensure that interactions which leak to these other states are not resonant in the dressed state basis. However, we assume that the other degenerate Rydberg states are detuned in some fashion, such as via another external drive. 
Furthermore, we assume the Rabi frequency for the polar molecules is sufficiently strong ($2 \pi \times 1$ MHz) that we can ignore any hyperfine structure of the bialkalis
\cite{suppcite}. The dressed states exhibit a diagonal interaction of the form
\begin{equation}
V(r,\theta) = \frac{C_3}{r^3} (1-3 \cos^2 \theta),
\end{equation}
where $C_3$ describes the strength of the dipolar interactions.

\begin{figure}
    \centering
    \includegraphics{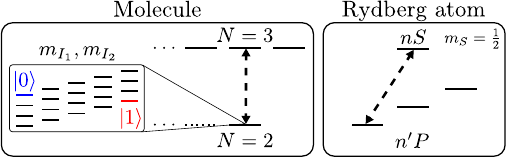}
    \caption{Level diagram for utilizing nuclear states in the presence of a magnetic field. Each rotational state hosts a nearly equivalent nuclear $m_{I_1}, m_{I_2}$ manifold, in which the qubit states are stored. The choice of nuclear states is largely arbitrary, although small deviations from identity interactions will vary slightly depending on the pair used due to hyperfine coupling \cite{suppcite}. The magnetic field allows us to tune the Rydberg transition frequencies into resonance with the molecule, after which either vdW or dipolar cooling may be utilized. Zeeman shifts on the rotational and nuclear states are much smaller than for the atom, so we neglect the rotational Zeeman shifts in the schematic.}
    \label{fig:hyperscheme}
\end{figure}

\emph{Hyperfine cooling}.---Finally, we discuss how the hyperfine structure from the nuclear spins $\hat{\mathbf{I}}_{i}$ of the two atoms ($i=1,2$) which form the molecules can readily realize the desired interactions. When the hyperfine structure is sufficiently small compared to other relevant energy scales, the molecule-Rydberg interactions are nearly independent of the nuclear $m_{I_1},m_{I_2}$ states and thus naturally realize the desired state-independent interactions within a given rotational state. 
This opens the possibility of tuning the transition frequencies into resonance with a magnetic field, dramatically expanding the possible Rydberg states which can be utilized, although we focus on S states (or dressed S $\to$ P states) for simplicity. We illustrate this in Fig.~\ref{fig:hyperscheme}, where we assume the angular momentum and spin of the Rydberg state are decoupled due to the large magnetic field.

An important consequence of this tunability is a change in the scaling of vdW interactions. This emerges because of the tunability of the atomic/molecular transition frequency difference $\delta_{\text{am}}$, which controls the strength of the vdW perturbation. By fixing $\delta_{\text{am}} \propto C_3/r_{\text{am}}^3$, the resulting vdW interactions become proportional to the dipolar interactions, providing an improvement in the cooling performance.
We determine $\delta_{\text{am}}$ by fixing the Rydberg pair-state to be composed of $95\,\%$ of $|rr\rangle$ at $r_{\text{am}}$.

\emph{Further Considerations}.---There are three additional considerations for the phonon exchange that must be accounted for: ensuring $G \ll \omega$ since for many state choices the coupling is too strong, the (anti)trapping of the Rydberg states, and Rydberg-Rydberg interactions, all of which can be addressed via off-resonant Rydberg dressing \cite{Pupillo2010,Johnson2010,Henkel2010, Jau2016,Zeiher2016,Zeiher2017, Arias2019, Borish2020,Guardado-Sanchez2021, Martin2021,Hines2023,Eckner2023,Weckesser2024}.
We weakly dress an atomic ground state $|g\rangle$ with Rydberg state $|r\rangle$ using Rabi frequency $\Omega_d$ and detuning $\Delta_d$  
to obtain the dressed  state $|d\rangle \approx |g\rangle + \frac{\Omega_d}{2 \Delta_d} |r\rangle$. By choosing a suitable Rydberg fraction $f \equiv \Omega_d^2/4\Delta_d^2$ and constraining the blockade radius to less than the lattice constant for the atoms, the bare phonon exchange rate, denoted $\mathcal{G}_{\text{am}}$, is rescaled to $G_{\text{am}} = f \mathcal{G}_{\text{am}}$.
Furthermore, the dressed Rydberg-Rydberg interactions, $\propto f^2$, can simultaneously be made suitably small so they do not interfere with the cooling in combination with suitable choices of the system geometry. Finally, by weakly dressing the ground state with the Rydberg state, we ensure that $|d\rangle$ experiences a trapping potential, which is crucial for phonon swap. 

Furthermore,  $|r\rangle$ is inherently dissipative due to Rydberg decay at rate $\gamma_r$, which results in an effective decay rate of $f \gamma_r$ for $|d\rangle$. When the decay occurs during phonon swap, not all phonons are exchanged. The swap efficiency is thus determined by the ratio $G_{\text{am}}/f \gamma_r = \mathcal{G}_{\text{am}}/\gamma_r$, which is independent of $f$,
and we find that a value $\mathcal{G}_{\text{am}}/\gamma_r \approx 15$ for a single atom-molecule pair results in an average of $95\,\%$ of the phonons being exchanged \cite{suppcite}. We define $r_{0.95}$ as the distance at which $\mathcal{G}_{\text{am}}/\gamma_r \approx 15$ is achieved.
Beyond $r_{0.95}$, phonon swap is no longer possible, 
and repumping of the dressed state is necessary, so it is better to utilize a combination of thermalization and continuous cooling of the auxiliary atoms.

We can thus investigate how the cooling performs as a function of $n$. There are two distance scales we are interested in: $r_{0.95}$ and half the LeRoy radius $a_\text{LR}/2$, i.e., the distance at which the Rydberg electron wavefunction begins to overlap with the molecule, which scales like $n^2$. Since $\mathcal{G}_{\text{am}}/\gamma_r \propto n^5/r^5, n^7/r^8$ for dipolar and vdW interactions, respectively, we find $r_{0.95} \propto n, n^{7/8}$, so higher $n$, and thus smaller $B_0$, are preferred if large separations are required. In the case of hyperfine cooling, the vdW interactions can scale like dipolar interactions, so $r_{0.95} \propto n$. If atom-molecule separations comparable to $a_\text{LR}/2$ are achievable (typically submicron scales), then $\mathcal{G}_{\text{am}}/\gamma_r \propto n^{-5}, n^{-9}$, so lower $n$, and thus larger $B_0$, are favorable. We find that the relevant LeRoy radii are on the order of half a micron, which in practice is very difficult to achieve, so we focus on $r_{0.95}$ as the most relevant figure of merit for the viability of phonon swap.

Finally, we comment on another natural approach: dc electric fields. In this case, no tuning of transition frequencies is needed since all states have an induced dipole moment. While recent experiments have demonstrated the viability of controllable interactions in strong electric fields \cite{Stecker2020}, we find that the phonon swap ranges are limited to small distances $r_{0.95} \lesssim \SI{0.75}{\micro\metre}$ and thus limited in their viability \cite{suppcite}, so we do not focus on this approach.

\begin{figure}
    \centering
    \includegraphics[scale=.5]{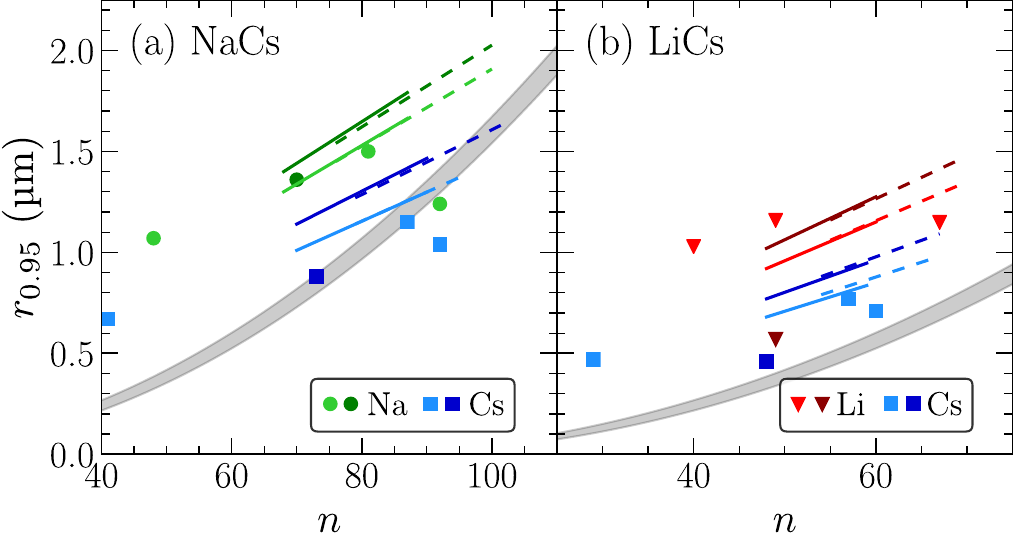}
    \caption{Phonon swap range for (a) NaCs and (b) LiCs when cooling with Na (green circles), Cs (blue squares), or Li (red triangles) via vdW (dark) or dipolar (light) interactions. The lines correspond to $r_{0.95}$ values achieved via tuning the magnetic field, with solid (dashed) lines denoting $N,m_N=2$ ($N,m_N=1$) molecular states. The shaded gray region indicates the range of LeRoy radii for the S, P, and D states \cite{Robertson2021}. Specific state and parameter values for each point can be found in the supplement \cite{suppcite}.}
    \label{fig:coolsum}
\end{figure}

\emph{Summary of results}.---We now summarize the performance of the cooling protocols for a selection of different Rydberg states \cite{Robertson2021}, molecular states, and magnetic fields in terms of the phonon swap range $r_{0.95}$. When utilizing a magnetic field, we consider three constraints. First, we assume a maximal field of 100 mT along the quantization axis. Second, we assume the Zeeman shift is at most of the order of the molecular transition. Finally, we restrict ourselves to only states where $r_{0.95}$ is larger than the LeRoy radius. The resulting values of $r_{0.95}$ are illustrated graphically in Fig.~\ref{fig:coolsum}.

As expected, the typical values of $n$ are higher for NaCs than those of LiCs due to the corresponding values of $B_0$ changing the resonance condition. In the absence of a magnetic field, the cooling tends to be limited to $r_{0.95} \lesssim \SI{1}{\micro\metre}$, and in some cases, $r_{0.95} < a_{\text{LR}}$, preventing phonon swap. One exception to this is the 70S state of Na, which exhibits a near-resonance with the NaCs $N=2 \to 3$ transition where $\delta_{\text{am}} \approx 2 \pi \times 6$ MHz, resulting in a relatively high $r_{0.95} \approx \SI{1.36}{\micro\metre}$. 

When a magnetic field is utilized with the nuclear spin states, the phonon swap range is enhanced and comparable to the above 70S Na near-resonance, and a range of higher $n$ can be utilized with larger $r_{0.95}$: $\SI{1}{\micro\metre} <r_{0.95}< \SI{2}{\micro\metre}$ for NaCs and $\SI{0.75}{\micro\metre} <r_{0.95}< \SI{1.5}{\micro\metre}$ for LiCs. Furthermore, this opens the possibility of using $N < 2$ states, whose smaller transition frequencies extend the possible Rydberg states to higher $n$. Remarkably, we find that $r_{0.95}$ is similar for vdW and dipolar interactions. This is a consequence of both the ability to tune $\delta_{\text{am}}$ for a given $r_{\text{am}}$ and because the interactions from several (virtual) transitions contribute to the vdW interaction while only a single transition contributes to the dipolar interaction.

\textit{Outlook}.---We have presented a method to achieve data-insensitive sympathetic cooling of molecules by Rydberg atoms, both through vdW interactions and dipolar interactions induced by microwave dressing. Although our focus was on utilizing molecule-Rydberg interactions for cooling, our work illustrates the potential capabilities of using external fields to control and tailor interactions in hybrid tweezer arrays of polar molecules and neutral atoms. Several future directions exist, including non-destructive molecular state measurement, enhanced molecule-molecule interactions via Rydberg atoms, or the transfer of quantum correlations from the atoms to the molecules or vice-versa.

\begin{acknowledgments}
We thank P.~Bienias, A.~M.~Kaufman, and A.~M.~Rey for helpful discussions.
J.T.Y.~was supported by the NWO Talent Programme (project number VI.Veni.222.312), which is (partly) financed by the Dutch Research Council (NWO). R.B.~acknowledges support by a Chicago Prize Postdoctoral Fellowship in Theoretical Quantum Science. A.V.G.~was supported in part by AFOSR MURI, DoE ASCR Quantum Testbed Pathfinder program (awards No.~DE-SC0019040 and No.~DE-SC0024220), NSF QLCI (award No.~OMA-2120757), NSF STAQ program, ARL (W911NF-24-2-0107), DARPA SAVaNT ADVENT, and NQVL:QSTD:Pilot:FTL. A.V.G.~also acknowledges support from the U.S.~Department of Energy, Office of Science, National Quantum Information Science Research Centers, Quantum Systems Accelerator and from the U.S.~Department of Energy, Office of Science, Accelerated Research in Quantum Computing, Fundamental Algorithmic Research toward Quantum Utility (FAR-Qu). K.-K.N.~acknowledges support from AFOSR MURI (FA9550-20-1-0323).
\end{acknowledgments}

\bibliography{MoleculeRydberg,Extra}

\end{document}


\title{Supplemental Material: Data-insensitive cooling of polar molecules with Rydberg atom}	

\author{Jeremy T. Young}
\email[Corresponding author: ]{j.t.young@uva.nl}
\affiliation{Institute of Physics, University of Amsterdam, 1098 XH Amsterdam, the Netherlands}

\author{Ron Belyansky}
\affiliation{Pritzker School of Molecular Engineering, University of Chicago, Chicago, IL 60637, USA}

\author{Kang-Kuen Ni}
\affiliation{Department of Chemistry and Chemical Biology, Harvard University, Cambridge, Massachusetts 02138, USA}
\affiliation{Department of Physics, Harvard University, Cambridge, Massachusetts 02138, USA}
\affiliation{Harvard-MIT Center for Ultracold Atoms, Cambridge, Massachusetts 02138, USA}

\author{Alexey V. Gorshkov}
\affiliation{Joint Quantum Institute and Joint Center for Quantum Information and Computer Science, NIST/University of Maryland, College Park, Maryland 20742 USA}

\date{\today}

\maketitle
\onecolumngrid

\renewcommand{\theequation}{S\arabic{equation}}

\renewcommand{\bibnumfmt}[1]{[S#1]}
\renewcommand{\citenumfont}[1]{S#1} 

\pagenumbering{arabic}

\makeatletter
\renewcommand{\thefigure}{S\@arabic\c@figure}
\renewcommand \thetable{S\@arabic\c@table}

This supplemental material is organized as follows: In Sec.~\ref{sec:phonon}, we derive the phonon exchange Hamiltonian. In Sec.~\ref{sec:chain}, we discuss how the phonon swap is modified by many-body effects in a 1D chain. In Sec.~\ref{sec:CaF}, we show how the cooling schemes can be adapted to the more complex level structure of CaF molecules.  
In Sec.~\ref{sec:vdW}, we present the derivations for the vdW cooling parameters. In Sec.~\ref{sec:dipolar}, we present the derivations for the dipolar cooling parameters. In Sec.~\ref{sec:hyper}, we discuss the role of hyperfine structure for the bialkali cooling schemes. In Sec.~\ref{sec:magnetic}, we present the details for the Zeeman splitting used for hyperfine cooling and the associated cooling parameters. In Sec.~\ref{sec:dc}, we present the derivations of cooling parameters for dc cooling. In Sec.~\ref{sec:fidelity}, we discuss effects of deviations from identity interactions on the fidelity.

\section{Phonon exchange Hamiltonian}
\label{sec:phonon}
In this section, we derive the phonon exchange Hamiltonian for a single atom and molecule pair with masses of $M_{\text{a/m}}$ and trapping frequencies $\omega_{\text{a/m}}$, respectively, which interact with the state-independent strength $V(\mathbf{r}_{\text{a}}-\mathbf{r}_{\text{m}})$, where $\mathbf{r}_{\text{a/m}}$ denote the positions of the corresponding particle. We assume that the position of each particle can be decomposed into a coherent (classical) part plus quantum fluctuations $\mathbf{r}_{\text{a/m}} \to \mathbf{r}_{\text{a/m}} + \hat{\mathbf{r}}_{\text{a/m}}$ and that the interactions are state-insensitive and can thus be captured purely by a distance-dependent potential $V$. The full Hamiltonian is
\begin{equation}
    \hat H = V(\delta \hat{\mathbf{r}}) + \sum_{\alpha =x,y,z} \sum_{\text{s} = \text{a},\text{m}} \left(\frac{\hat p_{s,\alpha}^2}{2 M_s} + \frac{M_s \omega_{s,\alpha}^2 \hat \alpha_{\text{s}}^2}{2} \right),
\end{equation}
where $\delta \hat{\mathbf{r}} \equiv \mathbf{r}_{\text{a}} + \hat{\mathbf{r}}_{\text{a}}- \mathbf{r}_{\text{m}}-\hat{\mathbf{r}}_{\text{m}} = (\mathbf{r}_{\text{a}} - \mathbf{r}_{\text{m}} ) + (\hat{\mathbf{r}}_{\text{a}}-\hat{\mathbf{r}}_{\text{m}})$, and $\hat p_{s,\alpha}$ is the momentum operator conjugate to the phonon position operator $\hat \alpha_{\text{s}}$, where $\text{s} \in \{\text{a},\text{m}\}$ denotes atomic or molecular phonons. We expand $V(\delta \hat{\mathbf{r}})$ around $\mathbf{r}_{\text{am}} \equiv \mathbf{r}_{\text{a}} - \mathbf{r}_{\text{m}}$ to second order in $\hat{\mathbf{r}}_{\text{a}} - \hat{\mathbf{r}}_{\text{m}}$, which results in the general equation
\begin{equation}
        V(\delta \hat{\mathbf{r}}) \approx \sum_\alpha (\hat \alpha_{\text{a}} - \hat \alpha_{\text{m}}) \partial_\alpha V(\mathbf{r}_{\text{am}}) + \frac{1}{2} \sum_{\alpha, \beta = x,y,z} (\hat \alpha_{\text{a}} - \hat \alpha_{\text{m}}) (\hat \beta_{\text{a}} - \hat \beta_{\text{m}}) \partial_\alpha \partial_\beta V(\mathbf{r}_{\text{am}}), \qquad \hat \alpha_{\text{s}} \equiv \frac{\hat s_\alpha + \hat s_\alpha^\dagger}{\sqrt{2 M_{\text{s}} \omega_{\text{s},\alpha}}}.
\end{equation}
Here, we note that the off-diagonal second-derivative terms in principle can couple phonons in different directions. However, if the corresponding trap frequencies $\omega_\alpha$ are different, as is often the case, these couplings will be off-resonant and can be ignored. Additionally, the linear terms will be zero when $V(\mathbf{r}_{\text{am}})$ is invariant under $\hat \alpha_{\text{a}} - \hat \alpha_{\text{m}} \to -(\hat \alpha_{\text{a}} - \hat \alpha_{\text{m}})$. The nonzero linear terms will be those in directions associated with the macroscopic separations. As shown by Ref.~\cite{Belyansky2019}, cooling in these directions can be achieved by smoothly varying the pulse used to turn the interactions on and off. This allows the particles to adiabatically follow the shifted equilibrium positions, thus always remaining in a minima with no net linear forces. In the case of continuous cooling, it is important that the cycling dynamics are sufficiently fast compared to the equilibration timescales of the particle positions.

Neglecting the terms which couple phonons of different directions, the remaining quadratic terms are
\begin{equation}
    \begin{aligned}
    \hat V_\alpha &\approx \frac{1}{2} \partial^2_\alpha V(\mathbf{r}_{\text{am}}) (\hat \alpha_{\text{a}} - \hat \alpha_{\text{m}})^2 = \frac{1}{2} \partial^2_\alpha V(\mathbf{r}_{\text{am}}) \left( \frac{\hat a_\alpha + \hat a_\alpha^\dagger }{\sqrt{2 M_{\text{a}} \omega_{\text{a},\alpha}}} - \frac{\hat m_\alpha + \hat m_\alpha^\dagger }{\sqrt{2 M_{\text{m}} \omega_{\text{m},\alpha}}} \right)^2 \\
    & \approx \frac{1}{2} \partial^2_\alpha V(\mathbf{r}_{\text{am}}) \left( \frac{(\hat a_\alpha + \hat a_\alpha^\dagger)^2}{2 M_{\text{a}} \omega_{\text{a},\alpha}} + \frac{(\hat m_\alpha + \hat m_\alpha^\dagger)^2}{2 M_{\text{m}} \omega_{\text{m},\alpha}} - 2\frac{(\hat a_\alpha + \hat a_\alpha^\dagger)(\hat m_\alpha + \hat m_\alpha^\dagger)}{\sqrt{4 M_{\text{a}} M_{\text{m}} \omega_{\text{a},\alpha}\omega_{\text{m},\alpha}}}\right),
    \end{aligned}
\end{equation}
and thus the phonon exchange coupling between molecules and atoms in the $\alpha$-direction is
\begin{equation}
    G_{\text{am},\alpha} \approx \frac{\partial_\alpha^2 V(\mathbf{r}_{\text{am}})}{\sqrt{4 M_{\text{a}} M_{\text{m}} \omega_{\text{a},\alpha}, \omega_{\text{m},\alpha}}},
\end{equation}
with similar expressions for the intraspecies phonon couplings as in the main text. The higher-order molecule-atom coupling terms fall off like $(M_{\text{a/m}} \omega_{\text{a/m}} r_{\text{am}}^2)^{-1/2}$. However, since the odd terms are suppressed via the rotating-wave approximation, the most relevant coupling is the quartic term, which is weaker than the quadratic by a factor $(\sqrt{M_1 M_2 \omega_{1} \omega_2 } r_{\text{am}}^2)^{-1}$, where the indices label the two new phonons in the product. In Table \ref{tab:mwr2}, we summarize the values these take at \SI{1}{\micro\metre} for the relevant scenarios considered in this work. We find the expansion to be well-justified if there are not too many phonons, although higher-order effects can become relevant more quickly for $^6$Li atoms.

\begin{table}
\centering
\def\arraystretch{1}
\setlength\tabcolsep{2.4mm}
\begin{tabular}{cccc}
\toprule
\toprule
 & $^{23}$Na & $^{133}$Cs & $^{23}$Na$^{133}$Cs\\
\midrule
$^{23}$Na & 0.018 & 0.0073 & 0.0068 \\
$^{133}$Cs & & 0.0030 & 0.0028\\
$^{23}$Na$^{133}$Cs & & & 0.0026 \\
 \bottomrule
 \bottomrule
\end{tabular}
\quad
\begin{tabular}{cccc}
\toprule
\toprule
 & $^{6}$Li & $^{133}$Cs & $^{6}$Li$^{133}$Cs\\
\midrule
$^{6}$Li & 0.067 & 0.014 & 0.014 \\
$^{133}$Cs & & 0.0030 & 0.0030 \\
$^{6}$Li$^{133}$Cs &  & & 0.0029\\
 \bottomrule
 \bottomrule
\end{tabular}

\vspace{.3cm}

\begin{tabular}{ccc}
\toprule
\toprule
 & $^{87}$Rb & $^{40}$Ca$^{19}$F \\
\midrule
$^{87}$Rb & 0.0047 & 0.0057  \\
$^{40}$Ca$^{19}$F  & & 0.0069  \\
 \bottomrule
 \bottomrule
\end{tabular}
\caption{Values of $(\sqrt{M_1 M_2  \omega_1 \omega
_2} r_{\text{am}}^2)^{-1}$ for all molecule, atom pairs considered in the main text and supplement. We assume $\omega_1 = \omega_2 = 2 \pi \times 25$ kHz and $r_{\text{am}} = \SI{1}{\micro\metre}$. \label{tab:mwr2}}
\end{table}

\subsection{Dissipation}

In the presence of dissipation, there is a nonzero probability that the Rydberg state decays prior to the swap time $t_{\text{swap}} \equiv \pi/2G_{\text{am}}$. Here, we calculate the expected fraction of molecular phonons remaining as a function of the decay probability $\gamma$.
The probability density function for a decay to take place at time $t$ is $\gamma e^{-\gamma t}$, where the coefficient $\gamma$ is due to the rate of dissipation, which is time-independent, and the factor $e^{-\gamma t}$ defines the probability the atom has not decayed yet. The expected fraction of molecular phonons after the swap is thus
\begin{equation}
    \int_0^{\pi/2 G_{\text{am}}} \gamma e^{- \gamma t} \cos^2( G_{\text{am}} t) dt = \frac{\gamma^2 + 2 G_{\text{am}}^2 \left(1 - e^{-\gamma t_{\text{swap}}} \right)}{\gamma^2 + 4 G_{\text{am}}^2},
\end{equation}
which is $0.05$ for $G_{\text{am}}/\gamma \approx 15.24$, corresponding to $r_{0.95}$ of the main text.

\section{Phonon-swap for 1D chain}
\label{sec:chain}
Here we generalize the results of the previous sections to the more realistic case of large register of atoms and molecules. For concreteness, we consider a 1D chain of $N$ auxiliary atoms (lattice constant $x_0$) that is brought to a distance of $y_0$ from a 1D chain of $N$ molecules.
For simplicity, we only consider a single trap direction ($z$) and the time-independent swap protocol. The generalizations to two or three directions and the adiabatic swap are straightforward \cite{Belyansky2019}.
The Hamiltonian for the vibrational modes in the $z$ direction is (we drop the $z$ labels from here on)
\begin{dmath}
	\label{eq:H-chain-full}
	\Hhat = \sum_{i=1}^N\omega_z (\ahat_{z,i}^\dagger\ahat_{z,i} + \mhat_{z,i}^\dagger\mhat_{z,i})-\frac{1}{2}\sum_{i=1,j=1,i\neq j}^N\qty[K_{ij}(\hat a_{z,i}-\hat a_{z,j})^2+L_{ij}(\hat m_{z,i}-\hat m_{z,j})^2]-\sum_{i=1, j=1}^NF_{ij}(\hat a_{z,i}-\hat m_{z,j})^2,
\end{dmath}
where by $\ahat_i\,(\mhat_i)$ we denote the phonon-annihilation operator for an atom (molecule) $i$ and by $\hat a_{z,i}=\frac{1}{\sqrt{2}}(\ahat+\ahat^\dagger)$ ($\hat m_{z,i}=\frac{1}{\sqrt{2}}(\mhat+\mhat^\dagger)$) the $z$ coordinate of an atom (molecule).
The $\frac{1}{2}$ is to avoid double-counting and the coefficients are given by
\begin{align}
		K_{ij} =\frac{G_{\text{aa}}}{\eta^{\alpha+2}\abs{i-j}^{\alpha+2}}\qc L_{ij} =\frac{G_{\text{mm}}}{\eta^{\alpha+2}\abs{i-j}^{\alpha+2}}\qc
		F_{ij} =\frac{G_{\text{am}}}{\qty[\eta^2(i-j)^2+1]^{(\alpha+2)/2}},
\end{align}
where $\alpha=3(6)$ for dipole-dipole (vdW) interactions and $\eta\equiv\frac{x_0}{y_0}$. To be consistent with the single atom-molecule pair case, we have defined $G$ with the nearest-neighbor separation between atoms and molecules ($y_0$). We have also assumed that particles that are farther apart than the nearest-neighbor separation (i.e., next nearest-neighbors and so on) experience power-law interactions. In other words, we assumed that the separation between next-nearest neighbors is significantly larger than any blockade radius due to potential weak Rydberg dressing and treat the interactions as pure power-laws. 

\begin{figure}[h!]
    \centering
    \includegraphics[width=0.5\linewidth]{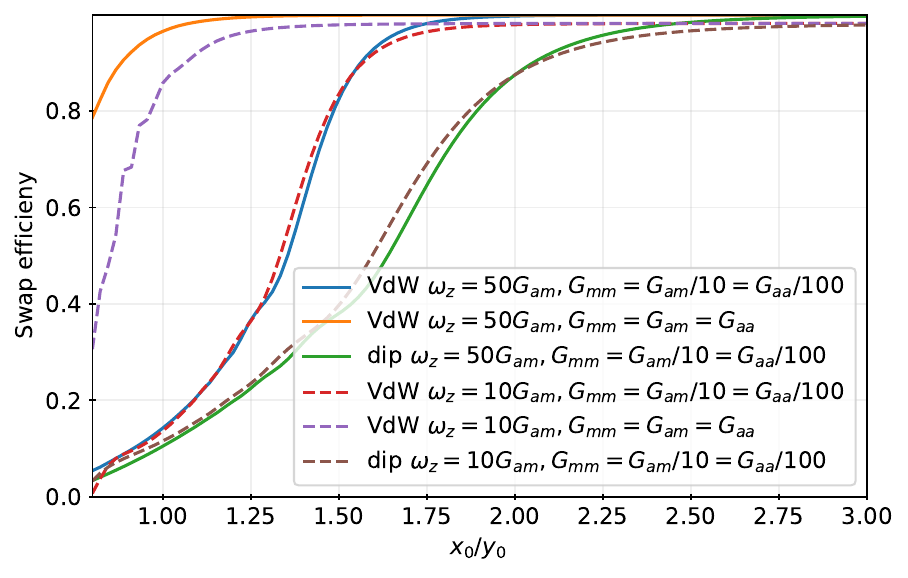}
    \caption{Phonon swap efficiency as a function of $x_0/y_0$ for several scenarios. We assume a chain of $N=50$ molecules and atoms. The atoms are initialized in their vibrational vacuum, while the molecules are initialized with $\expval*{\mhat_i^\dagger \mhat_i}=20$ for all $i$.}
    \label{fig:swapchain}
\end{figure}

\Cref{eq:H-chain-full} can be simulated exactly for large systems to determine the effectiveness of the phonon-swap in a large register of molecules and atoms. In Fig.~\ref{fig:swapchain}, we plot the swap efficiency, defined as $1-\bar{n}_m(\pi/2G_{\text{am}})/\bar{n}_m(0)$, where $\bar{n}_m(t) = \sum_{i=1}^N\expval*{\mhat_i^\dagger\mhat_i}(t)/N$ is the average phonon occupation in the molecules at time $t$, as a function of $x_0/y_0$ for several scenarios using exact numerics. First, we see that the swap efficiency approaches 1 more quickly for vdW interactions than for dipolar interactions, which is due to the vdW interactions approaching the limit of independent atom-molecule pairs more quickly. Second, we see that larger ratios $x_0/y_0$ are necessary when the atom-atom interactions are stronger than the molecule-atom interactions (when compared at equal distances), as this ensures that all interactions are comparable to or smaller than the nearest-neighbor atom-molecule interactions. Finally, we see that the swap efficiency is not strongly modified by reducing $\omega_z$ from $50 G_{\text{am}}$ to $10 G_{\text{am}}$, although the efficiency in the independent atom-molecule pair limit is reduced slightly further from 1 due to violations of the rotating-wave approximation  from finite $\omega_z$, with swap efficiencies of $0.999$ and $0.982$ in the cases of $\omega_z = 50 G_{\text{am}}$ and $\omega_z = 10 G_{\text{am}}$, respectively.

\section{C\lowercase{a}F Level Structure}
\label{sec:CaF}

\begin{figure}[h]
    \centering
    \includegraphics{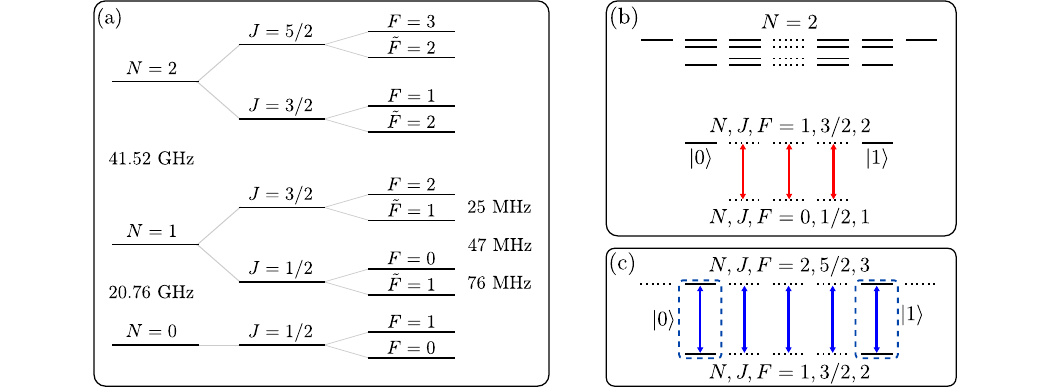}
    \caption{Level diagram and cooling schemes in the absence of a magnetic field for CaF. (a) Fine and hyperfine structure. Due to the strength of the hyperfine coupling, there is non-negligible mixing between manifolds with the same $N, F$, indicated by the tildes on the corresponding $F$ manifolds. (b) vdW cooling scheme. (c) Dipolar cooling scheme. The arrows indicate transitions which are driven with a microwave drive. The solid (dotted) levels correspond to states which are (not) involved in the vdW or dipolar interactions, both directly and virtually.}
    \label{fig:CaF}
\end{figure}

In this section, we discuss the hyperfine structure of $^{40}$Ca$^{19}$F, which is on the order of 10s of MHz, preventing the use of only rotational degrees of freedom. For the isotope of CaF under consideration, the ground state is a $^{2} \Sigma$ state, and the electron spin $S = 1/2$ couples to the molecular rotation $N$ to form $\mathbf{J} \equiv \mathbf{N} + \mathbf{S}$. Furthermore, the fluorine atom has nuclear spin of $I=1/2$, which couples to form the total angular momentum $\mathbf{F} = \mathbf{J} + \mathbf{I}$. The resulting level diagram is illustrated in Fig.~\ref{fig:CaF}(a). Here, we see that although the rotational constant $B_0$ is large compared to the bialkali atoms, we may readily utilize the $N=1, F=2$ states in a manner similar to the $N=2$ bialkali states, reducing the transition frequencies under consideration in comparison to the $N=2$ to $N=3$ transitions. 

The fine-structure and hyperfine Hamiltonians are
\begin{equation}
    H_{\text{fs}} = \gamma \mathbf{S} \cdot \mathbf{N}, \qquad H_{\text{hf}} = b_F \mathbf{S} \cdot \mathbf{I} + t_0 \sqrt{6} T^2(\mathbf{S}, \mathbf{I}) \cdot T^2(\mathbf{C}),
\end{equation}
where $\gamma = 2 \pi \times 39.49793(2)$ MHz describes the spin-rotation interaction, $b_F = 2 \pi \times 122.025(1)$ MHz the Fermi contact interaction, and $t_0 = 2 \pi \times 13.549(1)$ MHz the dipolar interactions \cite{Childs1981}.
There are no quadrupole interactions to consider. While the fine structure is diagonal in the $N,J,F$ basis, the hyperfine coupling mixes the states. Due to the large rotational energy scales, we neglect coupling between states with different $N$. Aside from the energy shifts, the hyperfine interaction mixes manifolds with equal $F$, such as $N=1,J=1/2,F=1$ and $N=1,J=3/2,F=1$.

The molecular level schemes we consider for vdW and dipolar cooling are illustrated in Figs.~\ref{fig:CaF}(b) and \ref{fig:CaF}(c), respectively. For vdW cooling, we see that there are now a variety of different $F,J$ in the $N=2$ manifold which will contribute to the vdW interactions of the $N,J,F=1,3/2,2, m_F = \pm 2$ states. As with the biakalis, a microwave drive must be utilized to prevent the molecules from leaving the qubit states due to resonant vdW interactions.
For dipolar cooling, we resonantly drive the same $N,J,F=1,3/2,2, m_F = \pm 2$ states with the $N,J,F=2,5/2,3, m_F = \pm 2$ states. Since the splitting is 10s of MHz, this does not strongly couple to the other $N=2$ manifolds. While the $N=2, J=2$ manifolds could be used instead, the strongest coupling is to the $F=3$ states.

\section{\lowercase{vd}W Cooling}
\label{sec:vdW}
In this section, we present the derivations of the vdW molecule-Rydberg interactions and their angular dependences. As in the main text, we assume the $|m_N| <2$ molecular states can be made non-degenerate with the $|m_N| = 2$ states through the application of a microwave drive, so we ignore interactions which change $m_N$. However, we will consider terms which do not conserve atomic $m_J$ as a means of determining their relevance.

Using the Wigner-Eckart theorem, the general expression for the vdW interactions with the molecule in the $|N, \pm N\rangle$ state and the Rydberg atom starting in the $|J,L,m_J\rangle$ state and ending in the $|J,L,m_J'\rangle$ state takes the form
\begin{subequations}
\begin{equation}
    V_{\text{vdW}}(r) = - \sum_p \frac{C_6^{(p)} \mathcal{D}^{(p)}}{r^6} ,
\end{equation}
\begin{equation}
    C_6^{(\tilde{L}\tilde{J};\tilde{N})} \equiv \sum_{n'} \frac{d_{N \tilde{N}} d_{\tilde{N} N} d_{nLJ, \tilde{n}\tilde{L}\tilde{J}} d_{\tilde{n}\tilde{L}\tilde{J}, nLJ}}{(E_N + E_{nLJ}) - (E_{\tilde{N}} + E_{\tilde{n}\tilde{L}\tilde{J}})},
\end{equation}
\begin{multline}
    \mathcal{D}^{(\tilde{J};\tilde{N})}_{m_J,m_J'} \equiv  |m_J\rangle \langle m_J'| \sum_{q_{\text{m}}, q_{\text{a}}} (-1)^{N +\tilde{N} \mp 2 N + q_{\text{m}} + J +  \tilde{J} -2 m_J + q_{\text{a}} } D_{q_{\text{a}}, q_{\text{m}}} D_{-q_{\text{a}} - \delta m_J, -q_{\text{m}}} \times \\
    \threej{N}{1}{\tilde{N}}{\mp N}{q_{\text{m}}}{\pm N - q_{\text{m}}} \threej{J}{1}{\tilde{J}}{-m_J}{q_{\text{a}}}{m_J - q_{\text{a}}} \times \\
    \threej{\tilde{N}}{1}{N}{\mp N + q_{\text{m}}}{-q_{\text{m}}}{\pm N} \threej{\tilde{J}}{1}{J}{q_{\text{a}}-m_J}{- q_{\text{a}} - \delta m_J}{m_J+\delta m_J},
\end{multline}
\begin{equation}
    D = \left( \begin{array}{ccc}
    -\frac{3}{2} e^{2 i \phi} \sin^2 \theta & - \frac{3}{\sqrt{2}}e^{i \phi} \cos \theta \sin \theta & \frac{1-3 \cos^2 \theta}{2} \\
    - \frac{3}{\sqrt{2}}e^{i \phi} \cos \theta \sin \theta & 1 - 3 \cos^2 \theta & \frac{3}{\sqrt{2}} e^{-i \phi} \cos \theta \sin \theta \\
    \frac{1-3 \cos^2 \theta}{2} & \frac{3}{\sqrt{2}} e^{-i \phi} \cos \theta \sin \theta & -\frac{3}{2} e^{-2 i \phi} \sin^2 \theta
    \end{array}
    \right),
\end{equation}
\end{subequations}
where $(p)$ indicates the channel for a given virtual process (i.e., defines the intermediate state manifold with quantum numbers $\tilde{N}$ for the molecule and $\tilde{L}, \tilde{J}$ for the atom), in which $C_6^{(p)}$ describes the overall scale of that corresponding interaction (in terms of the reduced dipole moments $d_{u, v} \equiv \langle u ||e r || v \rangle$ and energy differences $E$) and $\mathcal{D}^{(p)}$ is a matrix which captures the angular dependence of the interactions, where we define $\delta m_J \equiv m_J' - m_J$. The indices of $D$, which describes the angular dependence of the dipole-dipole interactions, are defined from left to right (top to bottom) as $-1, 0, +1$. Note that the $\tilde{L}$ dependence has been dropped from $(p)$ in $\mathcal{D}$ since it is independent of $\tilde{L}$. The matrices for the molecular states $|N,\pm N\rangle$ can be related by changing the sign of the various $m_J$ in the resulting matrix, taking the complex conjugate, and multiplying each entry by a factor $(-1)^{\delta m_J}$. We focus on the $|N,\pm N\rangle$ matrix except where otherwise noted. The reported values of the total $C_6$ are defined as $C_6 \equiv \sum_p C_6^{(p)} \mathcal{D}^{(p)}$ evaluated at $\theta = \pi/2$.

Assuming the Rydberg atom is in an S$_{1/2}$ state and the molecule in the $|2,2\rangle$ state,
we find for $m_J = \{-1/2,1/2\},m_J' = \{-1/2,+1/2\}$
\begin{equation}
    \mathcal{D}^{(1/2;3)} = \left(
    \begin{array}{cc}
    -\frac{95 - 51 \cos 2 \theta}{1260} & - \frac{ \sin 2 \theta}{30} e^{i \phi} \\ 
     - \frac{\sin 2 \theta}{30} e^{-i \phi}  & -\frac{67 + 33 \cos 2 \theta}{1260}
    \end{array}
    \right), 
    \qquad 
    \mathcal{D}^{(3/2;3)} = \left(
    \begin{array}{cc}
    \frac{37 + 6 \cos 2 \theta}{630} & - \frac{ \sin 2 \theta}{60} e^{i \phi} \\ 
     - \frac{\sin 2 \theta}{60} e^{-i \phi}  & \frac{44 - 15 \cos 2 \theta}{630}
    \end{array}
    \right).
\end{equation}
In the absence of fine structure splitting, the contribution from the P states together are equivalent to the contribution to S$_0$ ($L=0, J=0$) from the P$_1$ ($\tilde{L}=1, \tilde{J}=1$) channel
\begin{equation}
    \mathcal{D}^{(1;3)} = \frac{5 - \cos^2 \theta}{35},
\end{equation}
according to $-\frac{2}{3} \mathcal{D}^{(1/2;3)} + \frac{4}{3} \mathcal{D}^{(3/2;3)} = \mathcal{D}^{(1;3)} I_2$, where $I_2$ is a $2 \times 2$ identity matrix. The factors of $-2/3, 4/3$ come from the relation between the reduced dipole moments in the $J$ basis vs.~the $L$ basis for the Rydberg states. However, when the Rydberg fine structure is relevant, the interaction with the molecular $|2, \pm 2 \rangle$ states is no longer state-insensitive since there is no symmetry in the interactions when $m_J \to - m_J$ for $m_N$ fixed. This is captured by the fact that $\mathcal{D}^{(1/2;3)},\mathcal{D}^{(3/2;3)}$ are not individually proportional to identity. 

As in the main text, we approach this issue by using the symmetric superposition of the two. In this basis, the two matrices become
\begin{subequations}
\begin{equation}
    \tilde{\mathcal{D}}^{(1/2;3)} =
     - \frac{\mathcal{D}^{(1;3)} }{2} I_2 +
     2 \mathcal{M}, \qquad
    \tilde{\mathcal{D}}^{(3/2;3)} = \frac{\mathcal{D}^{(1;3)} }{2} I_2 + \mathcal{M}
    ,
\end{equation}
\begin{equation}
    \mathcal{M} = \left(
    \begin{array}{cc}
    -\frac{ \sin 2 \theta \cos \phi}{60} & \frac{1 - 3 \cos 2 \theta - 3i \sin 2 \theta \sin \phi}{180} \\ 
     \frac{1 - 3 \cos 2 \theta + 3i \sin 2 \theta \sin \phi}{180}  & \frac{\sin 2 \theta \cos \phi}{60}
    \end{array}
    \right),
\end{equation}
\end{subequations}
where $\mathcal{M}$ captures the effect of the asymmetry and the top left corresponds to the symmetric superposition $(|m_J=-1/2\rangle +|m_J = 1/2\rangle)/\sqrt{2}$ and the bottom right to the antisymmetric superposition $(|m_J=-1/2\rangle -|m_J = 1/2\rangle)/\sqrt{2}$. While we do not present it explicitly here, we find that this same structure holds for more complicated molecular hyperfine structures, such as utilized in CaF, with the caveat that the constant factors for $\mathcal{M}$ are channel-dependent and that the identity contribution is modified to the corresponding channel. 

Next, we consider the exchange interactions for $\mathcal{D}^{(1;3)}$, again taking the separation to be $r_{\text{am}}$ in the $y$-direction, from which we find
\begin{equation}
    \hat V_x^{(1;3)} \approx -\frac{3}{7} \frac{(\hat x_{\text{a}} - \hat x_{\text{m}})^2}{r_{\text{am}}^8}, \qquad 
    \hat V_y^{(1;3)} \approx -\frac{6}{7}\frac{\hat y_{\text{a}} - \hat y_{\text{m}}}{ r_{\text{am}}^7} + 3 \frac{(\hat y_{\text{a}} - \hat y_{\text{m}})^2}{ r_{\text{am}}^8}, \qquad 
    \hat V_z^{(1;3)} \approx -\frac{16}{35} \frac{(\hat z_{\text{a}} - \hat z_{\text{m}})^2}{r_{\text{am}}^8},
\end{equation}
which can be utilized for determining the contributions from $\mathcal{D}^{(1/2;3)},\mathcal{D}^{(3/2;3)}$. The remaining relevant contributions from $\mathcal{M}$ are all 0, except for a quadratic $x$-$z$ coupling term which we assume to be off-resonant as discussed in the derivation of the phonon exchange Hamiltonian. Thus the Rydberg fine structure matters only in how it modifies the $C_6$ coefficients. When reporting total $C_6$ values, we include the angular dependence at $\theta = \pi/2$, where it is also rotationally invariant in $\phi$. Similar calculations can be performed to identify the total perturbative pair-state mixing at first order in the wavefunction, which describes the validity of the perturbative expansion. We define the total fraction of the wavefunction from other pair-states as $f_{\text{mix}}$, which we likewise evaluate at $\theta = \pi/2$.

Similar expressions can be identified for CaF with the total angular momentum $F$ defining the molecular channel. Additionally, as noted above, the general structure of the interaction with the symmetrized Rydberg states holds, so we can neglect contributions from $\mathcal{M}$ since they do not affect the cooling for configurations with $\theta = \pi/2$. Noting that factors of $\pm 1/2$ must be taken into account for the Rydberg fine structure, we identify the relevant angular dependencies for $|F=2, m_F = \pm 2 \rangle$ to be
\begin{equation}
    \mathcal{D}^{(1;1)} = \frac{5 - 3 \cos^2 \theta}{30}, \qquad
    \mathcal{D}^{(1;2)} = -\frac{1 + \cos^2 \theta}{10}, \qquad
    \mathcal{D}^{(1;3)} = \frac{5 -  \cos^2 \theta}{35},
\end{equation}
although we note that $F, J, N$ will all have to be specified for the molecular channel in the equations for the $C_6^{(p)}$ coefficients. 
Here, we note that all terms are sums of constants with $\cos^2\theta$, so we consider the general interaction form $V(\hat r) = \frac{a+b \cos^2 \theta}{\hat r^6}$, which results in
\begin{equation}
    \hat V_x \approx - 3 a \frac{(\hat x_{\text{a}} - \hat x_{\text{m}})^2}{r_{\text{am}}^8} , \quad
    \hat V_y \approx -6 a\frac{(\hat y_{\text{a}} - \hat y_{\text{m}})}{r_{\text{am}}^7}  + 21 a \frac{(\hat y_{\text{a}} - \hat y_{\text{m}})^2}{r_{\text{am}}^8} , \quad 
    \hat V_z \approx (-3 a + b)\frac{(\hat z_{\text{a}} - \hat z_{\text{m}})^2}{r_{\text{am}}^8},
\end{equation}
where we have again assumed the separation to be $r_{\text{am}}$ in the $y$-direction.
Similar calculations can be carried out in the presence of additional fine and hyperfine structure, although the associated reduced dipole matrix elements must be accordingly updated.

In Table \ref{tab:vdW}, we present cooling parameters for vdW cooling. As discussed in the main text, we focus on Rydberg S states. For each atom, we include the three strongest $S$ state interactions. 49S for cooling LiCs with Cs is small due to the 49P$_{1/2}$ and 49P$_{3/2}$ transition frequencies being shifted from the molecular transition frequency in opposite directions, so their contributions to the vdW interactions have different signs, leading to a reduction in the overall interaction strength. Furthermore, we present values for an additional variable $f_{\text{mix}}$ at half the LeRoy radius. This defines the fraction of the molecule-atom pair state which is not composed of the associated bare states according to perturbation theory. Hence, it describes the extent to which the system is in the perturbative, vdW regime. For the validity of the scheme, this must be kept relatively small, which is typically the case except for near-resonances at short distances, such as the 70S state of Na for cooling NaCs or the 47S state of Rb cooling for CaF. 

\begin{table}[b]
\centering
\setlength\tabcolsep{1.92mm}
\begin{tabular}{cccccccccccc} \toprule \toprule
 & & $n$S & $C_6/2 \pi$ & $a_{\text{LR}}/2$ & $\mathcal{G}_{\text{am},z}/2 \pi$ & $\mathcal{G}_{\text{am}}/\gamma_r$ & $\mathcal{G}_{\text{am},z}/2 \pi$ & $\mathcal{G}_{\text{am}}/\gamma_r$ & $r_{0.95}$ & $f_{\text{mix}}$   \\
 & &  & (\SI{}{\kilo\Hz.\micro\metre^6}) & (\SI{}{\micro\metre}) & (kHz), [$a_{\text{LR}}/2$] & [$a_{\text{LR}}/2$] & (kHz), [\SI{1}{\micro\metre}] & [\SI{1}{\micro\metre}] & (\SI{}{\micro\metre}) & [$a_{\text{LR}}/2$] \\
\midrule
 & & 69S & $74.8$ & 0.77 & $13.7$ & 36.0 & 1.62 & 4.27 & 0.85 & $8.4 \times 10^{-4}$ \\
 & \ Na \ & \bf{70S} & $\bf{-3062}$ & \bf{0.79} & $\bf{442}$ & \bf{1218} & \bf{66.2} & \bf{182} & \bf{1.36} & \bf{1.5} \\
\multirow{2}{*}[-0.5\dimexpr \aboverulesep + \belowrulesep + \cmidrulewidth]{$^{23}$Na$^{133}$Cs} \ & & 71S & $-86.8$ & 0.81 & $9.94$ & 28.6 & 1.88 & 5.40 & 0.88 & $6.4 \times 10^{-4}$ \\
 \cmidrule{2-11}
 & & 72S & $71.4$ & 0.77 & $5.05$ & $12.0$ & 0.642 & 1.52 & 0.75 & $8.5 \times 10^{-4}$ \\
 & Cs & \bf{73S} & $\bf{-241}$ & \bf{0.80} & $\bf{13.5}$ & $\bf{33.4}$ & $\bf{2.16}$ & $\bf{5.36}$ & \bf{0.88} & $\bf{1.0 \times 10^{-2}}$ \\
 & & 74S & $61.3$ & 0.82 & $2.73$ & $7.04$ & $0.551$ & $1.42$ & 0.74 & $2.9 \times 10^{-3}$ \\
\midrule
 & & 47S & 1.38 & 0.36 & 203 & 110 & 0.0619 & 0.0334 & 0.47 & $1.8 \times 10^{-4}$ \\
 & Li & 48S & 5.4 & 0.38 & 566 & 325 & 0.242 & 0.139 & 0.56 & $1.6 \times 10^{-3}$ \\
 \multirow{2}{*}[-0.5\dimexpr \aboverulesep + \belowrulesep + \cmidrulewidth]{$^{6}$Li$^{133}$Cs} & & 
 \textbf{49S} & $\bf{-6.25}$ & \textbf{0.40} & \textbf{469} & \textbf{287} & \bf{0.280} & \bf{0.171} & \bf{0.57} & $\bf{1.2 \times 10^{-3}}$ \\
 \cmidrule{2-11}
 &  & 47S & 1.39 & 0.31 & 160 & 95.6 & 0.0132 & 0.00789 & 0.39 & $4.0 \times 10^{-4}$  \\
 & Cs & \textbf{48S} & \textbf{5.27} & \textbf{0.32} & \textbf{421} & \textbf{269} & \textbf{0.0502} & \bf{0.0321} & \bf{0.46} &  $\bf{4.0 \times 10^{-3}}$  \\
 & & 50S & $-3.36$ & 0.35 & 132 & 96.4 & 0.0320 & 0.0234 & 0.45 & $7.1 \times 10^{-4}$ \\
 \midrule 
 & & 46S & $-1.09$ & 0.31 & 235 & 146 & 0.0189 & 0.0117 & 0.41 & $6.4 \times 10^{-4}$ \\
 $^{40}$Ca$^{19}$F & $^{87}$Rb & \bf{47S} & $\bf{-31.5}$ & \bf{0.32} & \bf{4842} & \bf{3214} & \bf{0.560} & \bf{0.372} & \bf{0.63} & \bf{1.2} \\
  & & 48S & 0.921 & 0.34 & 96.2 & 68.4 & 0.0160 & 0.113 & 0.41 & $1.8 \times 10^{-4}$ \\
\bottomrule \bottomrule
\end{tabular}
\caption{Parameters for vdW cooling. Each row corresponds to cooling either $^{23}$Na$^{133}$Cs or $^6$Li$^{133}$Cs with the corresponding constituent Rydberg atom.  We assume $\omega_z = 2 \pi \times 25$ kHz. The entries which appear in Fig.~3 of the main text are in bold. Parentheses and square brackets denote units and the value of $r_{\text{am}}$ where the parameter was calculated, respectively, when applicable. \label{tab:vdW}}
\end{table}

\section{Dipolar Cooling}
\label{sec:dipolar}
\begin{table}[b]
\centering
\setlength\tabcolsep{.79mm}
\begin{tabular}{ccccccccccccc}
\toprule
\toprule
 &  & $nL \to n'L'$ & \ $m_J$ \  & $C_3/2 \pi$ & $a_{\text{LR}}/2$ & $\mathcal{G}_{\text{am}}/2 \pi$ & $\mathcal{G}_{\text{am}}/\gamma_r$ & $\mathcal{G}_{\text{am}}/2 \pi$ & $\mathcal{G}_{\text{am}}/\gamma_r$ & $r_{0.95}$ & $\Omega_{\text{max}}/\omega_{\text{m}}$ \\
 &  & &  & (\SI{}{\kilo\Hz.\micro\metre^3}) & (\SI{}{\micro\metre}) & (kHz), [$a_{\text{LR}}/2$] & [$a_{\text{LR}}/2$] & (kHz), [\SI{1}{\micro\metre}] & [\SI{1}{\micro\metre}] & (\SI{}{\micro\metre}) &  \\
 \midrule
 & & \bfseries \boldmath $48\text{P} \to 47\text{D}$ & \boldmath $1/2$ & \boldmath $568$ & \bfseries 0.39 & \boldmath $2.04 \times 10^3$ & \boldmath $2.54 \times 10^3$ & \bfseries  17.3 & \bfseries  21.5 & \bfseries  1.07 & \bfseries 0.21 \\
 & & $48\text{P} \to 47\text{D}$ & $3/2$ & $492$ & 0.39 & $1.76 \times 10^3$ & $2.21 \times 10^3$ & 14.9 & 18.7 & 1.04 & 0.17 \\
 \cmidrule{3-12}
 & \multirow{2}{*}{\ Na \ } & \bfseries \boldmath $81\text{P} \to 81\text{D}$ & \boldmath $1/2$ & \bfseries 674 & \bfseries 1.10 & \bfseries 12.9 & \bfseries 74.4 & \bfseries 20.5 & \bfseries 118 & \bfseries 1.50 & \bfseries 0.25 \\
 & & $81\text{P} \to 81\text{D}$ & $3/2$ & $584$ & 1.10 & 11.2 & 64.6 & 17.7 & 103 & 1.47 & 0.21 \\
 \cmidrule{3-12}
 & & \boldmath \bfseries $92 \text{P} \to 90 \text{D}$ & \bfseries 1/2 & \bfseries  158 & \bfseries 1.42 & \bfseries 0.844 & \bfseries 7.83 & \bfseries 4.80 & \bfseries 44.5 & \bfseries 1.24 & \bfseries 0.06 \\
 \multirow{2}{*}[-0.5\dimexpr \aboverulesep + \belowrulesep + \cmidrulewidth]{$^{23}$Na$^{133}$Cs \ } & & $92 \text{P} \to 90 \text{D}$ &3/2 & 136 & 1.42 & 0.728 & 6.85 & 4.14 & 39.0 & 1.21 & 0.048 \\
 \cmidrule{2-12}
 & & \bfseries \boldmath $41\text{P} \to 40\text{D}$ & \bfseries 1/2 & \bfseries 189 & \bfseries 0.25 & \boldmath $2.54 \times 10^3$ & \boldmath $2.22 \times 10^3$ & \bfseries 2.39 & \bfseries 2.09 & \bfseries 0.67 & \bfseries 0.14 \\
 & &  $41\text{P} \to 40\text{D}$ & $3/2$ & $295$ & 0.25 & $3.96 \times 10^3$ & $1.78 \times 10^3$ & 3.73 & 1.67 & 0.64 & 0.11 \\
 \cmidrule{3-12}
 & \multirow{2}{*}{Cs} & \bfseries \boldmath $87\text{P} \to 85 \text{D}$ & \bfseries 1/2 & \bfseries 566 & \bfseries 1.20 & \bfseries 2.93 & \bfseries 12.5 & \bfseries 7.15 & \bfseries 30.6 & \bfseries 1.15 & \bfseries 0.22 \\
 & & $87\text{P} \to 85 \text{D}$ & 3/2 & $-476$ & 1.20 & 2.46 & 10.13 & 6.02 & 24.7 & 1.10 & 0.17\\
 \cmidrule{3-12}
 & & \bfseries \boldmath $92\text{P} \to 92\text{D}$ & \bfseries 1/2 & \bfseries $-282$ & \bfseries 1.34 & \bfseries 0.821 & \bfseries 4.28 & \bfseries 3.56 & \bfseries 18.5 & \bfseries 1.04 & \bfseries 0.11 \\
 & & $91\text{P} \to 91\text{D}$ & $3/2$ & $249$ & 1.31 & 0.813 & 3.94 & 3.15 & 15.3 & 1.00 & 0.087 \\
 \midrule
 & & $39\text{S} \to 39\text{P}$ & $1/2$ & $-429$ & 0.25 & $25.6 \times 10^3$ & $13.8 \times 10^3$ & 27.0 & 14.6 & 0.99 & 0.03 \\
 &  & \bfseries \boldmath $40\text{S} \to 40 \text{P}$ & \bfseries 1/2 & \bfseries 368 & \bfseries 0.27 & \boldmath $1.70 \times 10^{4}$ & \boldmath $13.0 \times 10^3$ & \bfseries 231 & \bfseries 17.7 & \bfseries 1.03 & \bfseries 0.031 \\
 \cmidrule{3-12}
 & \multirow{2}{*}{Li} & $48\text{S} \to 47\text{P}$ & $1/2$ & $401$ & 0.39 & $2.99 \times 10^3$ & $33.7 \times 10^3$ & 25.2 & 28.4 & 1.13 & $0.03$ \\
 \multirow{2}{*}[-0.5\dimexpr \aboverulesep + \belowrulesep + \cmidrulewidth]{$^{6}$Li$^{133}$Cs} & & \boldmath \bfseries $49\text{S} \to 48\text{P}$ & \bfseries 1/2 & \bfseries $-418$ & \bfseries 0.40 & \boldmath $2.54 \times 10^3$ & \boldmath $3.04 \times 10^3$ & \bfseries 26.4 & \bfseries 31.5 & \bfseries 1.16 & \boldmath $0.03$  \\
 \cmidrule{3-12}
 & & $66 \text{S} \to 64 \text{P}$ & 1/2 & $134$ & 0.73 & 41.2 & 139 & 8.43 & 28.5 & 1.13 & 0.011 \\
  & & \bfseries \boldmath $67 \text{S} \to 65 \text{P}$ & \bfseries 1/2 & \boldmath $-125$ & \bfseries 0.75 & \bfseries 33.2 & \bfseries 126 & \bfseries 7.89 & \bfseries 30.0 & \bfseries 1.15 & \bfseries 0.011 \\
 \cmidrule{2-12}
 & & \bfseries \boldmath $29\text{P} \to 28\text{D}$ & \bfseries 1/2 & \bfseries $-92.4$ & \bfseries 0.12 & \boldmath $55.2 \times 10^3$ & \boldmath $15.0 \times 10^3$ & \bfseries 1.24 & \bfseries 0.337 & \bfseries 0.47 & \bfseries 0.015  \\
 &  & $29\text{P} \to 28 \text{D}$ & 3/2 & $-69.5$ & 0.12 & $41.5 \times 10^3$ & $11.9 \times 10^3$ & 0.931 & 0.266 & 0.45 & 0.012 \\
 \cmidrule{3-12}
 & \multirow{2}{*}{Cs} & \bfseries \boldmath  $57 \text{P} \to 55 \text{D}$ & \bfseries 1/2 & \boldmath $-202$ & \bfseries 0.50 &  \bfseries 88.7 & \bfseries 149 &  \bfseries 2.70 &  \bfseries 4.55 &  \bfseries 0.77 & \bfseries 0.020 \\ 
 & & $57 \text{P} \to 55 \text{D}$ & 3/2 & $-171$ & 0.50 & 75.2 & 134 & 2.29 & 4.09 & 0.77 & 0.016\\ 
 \cmidrule{3-12}
 & & \bfseries \boldmath $60\text{P} \to 60 \text{D}$ & \bfseries 1/2 & \bfseries $-126$ & \bfseries 0.55 & \bfseries 32.6 & \bfseries 51.7 & \bfseries 1.69 & \bfseries 2.69 & \bfseries 0.71 & \bfseries 0.011 \\
 & & $60\text{P} \to 60 \text{D}$ & 3/2 & $-111$ & 0.55 & 28.7 & 47.8 & 1.49 & 2.49 & 0.70 & 0.0088 \\
 \midrule
 & & $39\text{P} \to 38 \text{D}$ & 1/2 & $-184$ & 0.24 & $6.27 \times 10^3$ & $3.50 \times 10^3$ & 4.69 & 2.62 & 0.70 & 0.048 \\
 & & $39\text{P} \to 38 \text{D}$ & 3/2 & $-162$ & 0.24 & $5.51 \times 10^3$ & $3.14 \times 10^3$ & 4.12 & 2.35 & 0.68 & 0.040 \\
 \cmidrule{3-12}
 \multirow{2}{*}{$^{40}$Ca$^{19}$F} & \multirow{2}{*}{$^{87}$Rb} & $51\text{P} \to 49 \text{D}$ & 1/2 & $190$ & 0.39 & 35.3 & 80.8 & 0.335 & 0.766 & 0.55 & 0.033 \\ 
 & & $51\text{P} \to 49 \text{D}$ & 3/2 & $-11.3$ & 0.39 & 24.5 & 51.0 & 0.233 & 0.484 & 0.50 & 0.026 \\
 \cmidrule{3-12}
 & & $61\text{P} \to 61 \text{D}$ & 1/2 & 3.51 & 0.60 & 1.19 & 2.67 & 0.0892 & 0.200 & 0.42 & 0.0013\\
 & & $61\text{P} \to 61 \text{D}$ & 3/2 & 2.98 & 0.60 & 1.02 & 2.33 & 0.0759 & 0.174 & 0.41 & 0.0011 \\
\bottomrule
\end{tabular}
\caption{Optimal parameters for dipolar cooling. 
A drive strength of $2 \pi \times 1$ MHz is assumed for the molecules, and $\Omega_{\text{max}}/\omega_{\text{m}}$ denotes the extent to which this drive is perturbative. We assume $\omega_z = 2 \pi \times 25$ kHz. The entries which appear in Fig.~3 of the main text are in bold. Parentheses and square brackets denote units and the value of $r_{\text{am}}$ where the parameter was calculated, respectively, when applicable. \label{tab:dd}}
\end{table}

In this section, we discuss the general form of the phonon exchange rates in the case of dipolar interactions, which will apply to the approaches which use microwave dressing or the application of an electric field. We assume that the macroscopic separation $r_{\text{am}} = |\mathbf{r}_{\text{am}}| = |\mathbf{r}_{\text{a}} - \mathbf{r}_{\text{m}}|$ is in the $y$-direction and that $V(\mathbf{r}_{\text{am}}) = C_3 (1- 3 \cos^2 \theta_{\text{am}})/r^3$, corresponding to a dipolar interaction cooling scheme. If this relative separation is changed, some of the following analysis will be modified as a result of the angular dependence of the interactions. The interaction terms are
\begin{subequations}
    \begin{equation}
        \hat V_x \approx -\frac{3}{2}\frac{(\hat x_{\text{a}} - \hat x_{\text{m}})^2}{ r_{\text{am}}^5}, \qquad
        \hat V_y \approx -3\frac{\hat y_{\text{a}} - \hat y_{\text{m}}}{r_{\text{am}}^4} + 6 \frac{(\hat y_{\text{a}} - \hat y_{\text{m}})^2}{r_{\text{am}}^5}, \qquad 
        \hat V_z \approx -\frac{9}{2}\frac{(\hat z_{\text{a}} - \hat z_{\text{m}})^2}{r_{\text{am}}^5}.
    \end{equation}
\end{subequations}
In the above equations, we see that a linear term is present in the $y$-direction since this is the direction of macroscopic separation. In Table \ref{tab:dd}, we present cooling parameters for dipolar cooling. As discussed in the main text, for Na, Cs, and Rb, we dress $\text{P}\to\text{D}$ transitions, while for Li we dress $\text{S}\to\text{P}$ transitions. We present the behavior for several near-resonant transitions and choices of $m_J$ for each atom/molecule pair. 

\section{Hyperfine structure}

\label{sec:hyper}
In this section, we discuss the effect of hyperfine structure on the bialkali molecule interactions. While we focus on hyperfine coupling, the effects of other sources of coupling and energy shifts on the structure of the identity interactions, such as weak external fields and vectors shifts, can be understood in a similar fashion. As discussed in the main text, in the absence of hyperfine coupling, the nuclear spin degrees of freedom naturally provide a way for realizing data-insensitive interactions without relying on the $m_N \to -m_N$ symmetry or $\delta m_N > 2$. This is a consequence of the fact that without hyperfine couplings, the interactions with the Rydberg atom within a given rotational state are proportional to identity in the nuclear spin basis. However, in the presence of hyperfine coupling, there will be deviations from identity interactions.

The hyperfine Hamiltonian takes the form
\begin{subequations}
\begin{equation}
    H_{\text{hf}} = H_Q + H_{IN} + H_{\text{t}} + H_{\text{sc}},
\end{equation}
\begin{equation}
    H_Q = -e \sum_{i=1}^2 T^2(\nabla \mathbf{E}_i) \cdot T^2(\mathbf{Q}_i), \qquad H_{IN} = \sum_{i = 1}^2 c_i \mathbf{N} \cdot \mathbf{I}_i, \qquad H_{\text{t}} = -c_3 \sqrt{6} T^2(\mathbf{C}) \cdot T^2(\mathbf{I}_1, \mathbf{I}_2), \qquad H_{\text{sc}} = c_4 \mathbf{I}_1 \cdot \mathbf{I}_2,
\end{equation}
\end{subequations}
where $\mathbf{I}$ are the nuclear spins of the two alkali atoms, $e$ is the electric charge, $T^k_p$ denotes the $p$th component of a rank $k$ tensor and $T^k_p(\mathbf{C}) \equiv C_p^k(\theta,\phi) = \sqrt{4 \pi/(2k+1)} Y_{k,p}(\theta,\pi)$ is a (renormalized) spherical harmonic. The constants associated with each term are summarized in Table \ref{tab:molecules}.
Here, $H_Q$ denotes the quadrupole interaction, which can be recast in the form $H_Q = \sum_{i=1}^2 (eQq)_i \frac{\sqrt{6}}{4 I_i (2 I_i -1)} T^2 (\mathbf{C}) \cdot T^2(\mathbf{I}_i, \mathbf{I}_i) $, where $eQ_i$ is the electric quadrupole moment and $q_i$ the negative of the electric field gradient of nucleus $i$. $H_{IN}$ denotes the spin-rotation coupling, while $H_{\text{t}}$ and $H_{\text{sc}}$ denote the tensor and scalar interactions between the nuclear dipole moments. 

Of these, $H_Q, H_{IN}, H_{\text{t}}$ can couple rotational states with different $m_N$. To mitigate this coupling, we must ensure that states with different $m_N$ are non-degenerate. For vdW cooling, this is achieved through the additional dressing field that couples the $N=1$ states to the $N=2$ states. Given the hyperfine coupling is on the order of 10s of kHz, we assume the drive strength, and thus the dressed energies, is on the order of MHz, mitigating these processes. For the dipolar cooling, the drive itself breaks this degeneracy for both $N=2$ and $N=3$ states, which can likewise be on the order of MHz. For $^{23}$Na$^{133}$Cs and $^{6}$Li$^{133}$Cs, the hyperfine coupling is dominated by the electric-quadrupole interaction, where an uncoupled basis (in terms of the two nuclear spins) is a good basis. For $H_Q$ and $H_{IN}$, the individual magnetic quantum numbers for $\mathbf{I}_1,\mathbf{I}_2$ $(m_{I_1},m_{I_2})$ are conserved when $m_N$ is conserved, so they only determine the energy shifts in the uncoupled basis. In contrast, $H_{\text{t,sc}}$ only conserve total $m_I$, so they perturbatively mix the hyperfine states. There is an exception for the stretched states, which will not mix and can thus be directly used.

\begin{table}[b]
\centering
\def\arraystretch{1}
\setlength\tabcolsep{2.4mm}
\begin{tabular}{ccccccccc}
\toprule
\toprule
   & $I_1$ & $I_2$ & $(eQq)_1/2 \pi$ (MHz) & $(eQq)_2/2 \pi$ (MHz) & $c_1/2 \pi$ (Hz) & $c_2/2 \pi$ (Hz) & $c_3/2 \pi$ (Hz) & $c_4/2 \pi$ (Hz)\\
\midrule
 $^{23}$Na$^{133}$Cs & 3/2 & 7/2 & $-0.097$ & 0.150 & 14.2 & 854.5 & 105.6 & 3941.8 \\
 $^{6}$Li$^{133}$Cs & 1 & 7/2 & $3 \times 10^{-4}$ & 0.181 & 15.2 & 3475.5 & 53.1 & 620.8 \\
 \bottomrule
 \bottomrule
\end{tabular}
\caption{Molecular hyperfine splitting parameters taken from Ref.~\cite{Aldegunde2017}. \label{tab:molecules}}
\end{table}

First, we discuss how the identity vdW interactions are modified. There are two ways in which the hyperfine interactions play a role: (i) They modify the energies in the denominator of the perturbative expansion. In general, we can neglect this contribution. Even for the smallest pair-state energy difference of 70S Na with NaCs of approximately $2 \pi \times 6$ MHz (the others are typically 10s to 100s times larger), the hyperfine structure is on the order of 10s of kHz. Hence, the deviation can be estimated as no more than $1\,\%$, which can be mitigated even further by choosing nuclear states with equal shifts. (ii) They mix the nuclear states. In this case, this mixture occurs on the level of the $N=2$ states and the intermediate $N=3$ states. However, neither will enhance the deviations that arise from the hyperfine energy shifts aside from modifying where the deviations are. To understand this, we note that if the interactions are written in the $|N,m_N\rangle$ basis, then the rotations in the nuclear $N=2$ states commute with the interaction terms in the perturbative expansion due to the extra drive, only entering via a rotation of the inverse energy difference. Likewise, the rotations in the nuclear $N=3$ states (which can mix the degenerate $m_N$) do not affect the $N=2$ space, so the effect is also only on the inverse energy difference. 

For strong magnetic fields, the nuclear states decouple, and there is minimal mixing, so the hyperfine Hamiltonian only shifts the energy, leaving the interaction matrix fully diagonal in terms of the nuclear states. In this regime, the Rydberg and molecule interactions can be brought close enough to resonance that these energy shifts could become important. In this case, one should choose pairs of states whose magnetic quantum numbers are the same (in magnitude) for each nuclear spin, which will ensure that the hyperfine energy shifts are equal.

For dipolar cooling, again the energy shifts play the most important role in modifications from the identity. In the absence of these energy shifts, each nuclear hyperfine eigenstate in the $N=2$ manifold is driven with the same Rabi frequency to an equivalent superposition of the $N=3$ nuclear states. However, since the eigenstates are not equivalent within the two hyperfine manifolds, once the energy shifts are incorporated, the $N=2$ states are driven to a superposition of hyperfine eigenstates of the $N=3$ manifold which have different energies and thus different detunings, leading to non-identity interactions and leakage to different nuclear states via the interactions. To minimize this, the drive must be large compared to the detunings. The deviation from identity interactions will be approximately the square of the ratio of half the variance of the detunings to the Rabi frequency, which we define as $\epsilon^2$. 

We define the microwave-dressed rotational states $|\pm,m_N\rangle = (|N=2,m_N\rangle \pm |N=3,m_N\rangle)/\sqrt{2})$ and choose the symmetric superpositions $|+,m_N,m_{I_1},m_{I_2}\rangle$ for the molecular qubit. Due to the energy shifts, the hyperfine structure will dress these with the antisymmetric superpositions $|-,m_N, m_{I_1}',m_{I_2}'\rangle$ with coefficients proportional to $\epsilon$, which defines our new, perturbed dressed hyperfine manifold. The modifications to the interactions are thus described by the modifications to the corresponding dipole moments of these states since the microwave-dressed Rydberg state is unchanged. Since there are no transition dipole matrix elements between $|+,m_N\rangle$ and $|-,m_N\rangle$, the only modifications to the perturbed manifold are due to the permanent dipole moments of $|\pm,m_N\rangle$, so these shifts will be proportional to $\epsilon^2$ as a result. This emphasizes further the importance of using resonant microwave dressing on the molecule rather than the Rydberg atom. While the coupling between the pair of perturbed $|\pm,m_N\rangle$ manifolds are of order $\epsilon$, these are highly off-resonant due to the dressed energies. For both molecules, the full range of detunings is on the order of $2 \pi \times 100$ kHz, so a drive of $2 \pi \times 1$ MHz limits the deviations from the identity matrix and the corresponding leakage processes to less than approximately $0.25\,\%$. 

\section{Hyperfine cooling}
\label{sec:magnetic}
In this section, we discuss the details of the application of a magnetic field for tuning the cooling schemes using the hyperfine structure.
First, we discuss the effects of the magnetic field on the level structures of the molecule and atom. Second, we discuss how the use of a magnetic field changes the angular dependence of the resulting vdW interactions.

\subsection{Level Structure}

The molecular Zeeman Hamiltonian is given by
\begin{equation}
    H_{\text{Z},m} = g_s \mu_B \mathbf{S} \cdot \mathbf{B} - g_r \mu_{\text{N}}  \mathbf{N} \cdot \mathbf{B} - \sum_{i=1}^2 g_i \mu_{\text{N}}  \mathbf{I}_i \cdot \mathbf{B} (1-\sigma_i),
\end{equation}
where $\mu_{\text{B}}$ ($\mu_{\text{N}}$) is the Bohr (nuclear) magneton, $g_s \approx 2$ is the Land\'{e} g-factor for the electron spin, $g_r$ is the Land\'{e} g-factor of rotations, and the $g_i$ are the Land\'{e} g-factors for the two nuclei, with $\sigma_i$ defining their shielding factors. Generally, $g_r$ is a hundred times smaller than $g_i$, so we neglect it. The first term for the spin degree of freedom only applies to CaF and always dominates when present. Values for the different coefficients are shown in Table \ref{tab:moleculesB}. The atomic Zeeman Hamiltonian is given by
\begin{equation}
    H_{\text{Z},a} = \mu_{\text{B}} \mathbf{B} \cdot (g_L \mathbf{L} + g_s \mathbf{S}),
\end{equation}
where $g_L = 1$ is the Land\'{e} g-factor for the angular momentum. 

\begin{table}
\centering
\def\arraystretch{1}
\setlength\tabcolsep{2.4mm}
\begin{tabular}{ccccc}
\toprule
\toprule
   & $g_1$ & $g_2$ & $\sigma_1$ & $\sigma_2$ \\
\midrule
 $^{23}$Na$^{133}$Cs & 1.478 & 0.738 & $6.392 \times 10^{-4}$ & $6.2787 \times 10^{-3}$ \\
 $^{6}$Li$^{133}$Cs & 0.822 & 0.738 & $1.080 \times 10^{-4}$ & $6.2440 \times 10^{-3}$ \\
 $^{40}$Ca$^{19}$F & 0 & 5.585 & 0 & - \\
 \bottomrule
 \bottomrule
\end{tabular}
\caption{Molecular hyperfine Zeeman parameters taken from Ref.~\cite{Aldegunde2017} for bialkalis and from Ref.~\cite{Anderegg2019a} for CaF. For CaF, the shielding factor $\sigma_2$ is neglected since the shift from $g_2$ is already small compared to the shift due to the electron spin. 
\label{tab:moleculesB}}
\end{table}

\begin{figure}[b]
    \centering
    \includegraphics{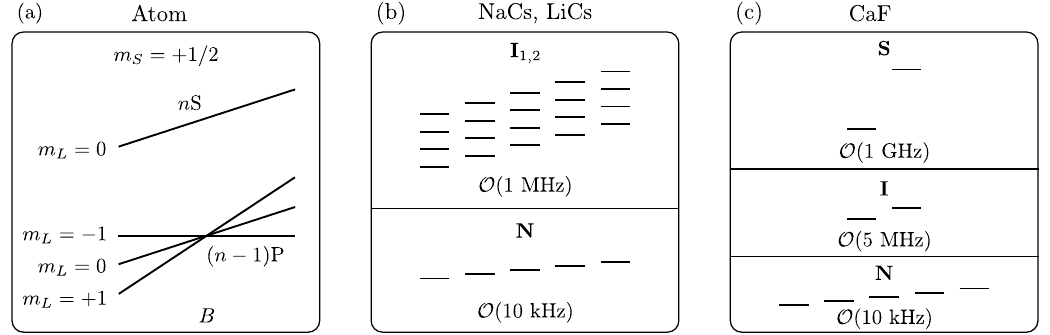}
    \caption{Zeeman splitting of (a) atoms, (b) bialkalis, and (c) CaF at $B=100$ mT. (a) Depending on the sign of $B$, choice of $m_S$, and relative energy of the S and P states, the stretched state and the $m_L = - 2 m_S$ state will be the states with largest/smallest transition frequencies. There is (almost) no shift for the $m_L = 0$ state since $g_s \approx 2 g_L$}
    \label{fig:dressB}
\end{figure}

Since the ratio of the Bohr and nuclear magnetons is $\mu_{\text{B}}/\mu_{\text{N}} \approx 1836$, the resonance tuning is dominated by Rydberg Zeeman shifts. While CaF has a strong Zeeman shift due to the electron spin, at strong fields it enforces conservation of $m_S$, so there is no tunability of the transition frequency. For the molecule, the Zeeman shift's role is primarily to mitigate the hyperfine coupling. In terms of energy shifts, $\mu_{\text{B}} \approx 2 \pi \times 14$ MHz/mT, while $\mu_{\text{N}} \approx 2 \pi \times 7.6$ kHz/mT. At magnetic fields of 100 mT, this corresponds to Zeeman splittings (between neighboring $m_I$ states) on the order of $2 \pi \times 1.4$ GHz for the atoms (or electron spin) and $2\pi \times 0.76$ MHz for the molecules (minus electron spin), up to the various g-factors. Approximate scales at 100 mT are shown in Fig.~\ref{fig:dressB}.

Next, we discuss the choice of qubit states. Since hyperfine nuclear states readily realize (near) identity interactions, the protocol is no longer constricted by the same symmetry considerations as for purely rotational degrees of freedom. For example, the $N=0$ or $N=1$ states can be used, which will allow for smaller transition frequencies. Note, however, that resonant vdW flip-flop interactions between different $m_N$ states must still be considered since the corresponding Zeeman shifts are small (on the order of 10s of kHz). Hence, additional drives must still be applied to shift the rotational states which are not involved in the protocol. 

For CaF, the electronic spin degree of freedom will quickly decouple from the other degrees of freedom as the magnetic field is turned on. Two of the options for encoding the qubit are the electronic spin (which may then need to be recoupled by exciting, for example, other electronic states) and the nuclear spin. 
Here we focus on using the nuclear spin states $m_I = \pm 1/2$ with fixed electronic spin projection $m_S$, like for the bialkalis. 
Provided $|N,m_N\rangle \to |N,m_N'\rangle$ transitions are forbidden via external drives, in the strong magnetic field limit where $m_S$ is conserved, we may rewrite $H_{\text{fs}} \approx \gamma S_z N_z$ and $H_{\text{hf}} \approx b_F S_z I_z + 2 t_0 I_z S_z \mathbf{C}_0^2$, where all off-diagonal terms have been dropped as off-resonant.

For the bialkalis, this is more complicated due to the presence of two nuclear spins. Since $m_{I_1} + m_{I_2}$ is conserved, the off-diagonal contributions in the hyperfine structure must be compared to the energy scales set by $g_1 - g_2$ (plus shielding factors) rather than the two individually. This can be potentially problematic for molecules like $^6$Li$^{133}$Cs, where these are relatively close, reducing the Zeeman splitting scales by roughly a factor of 10. However, since $c_3,c_4$, which give rise to mixing the nuclear states, are also relatively small, the reduced splitting is less of an issue. Nevertheless, for a specific choice of magnetic field it is important to carefully consider this effect. As in the absence of a magnetic field, the utilization of the stretched states can avoid this issue.

Since the molecule and atom are brought into resonance at large fields, the shifts from the quadratic Zeeman effect are in principle relevant since they are on the order of MHz.
However, this will primarily be important in determining what magnetic field to use for a given relative transition frequency. While higher-order Zeeman shifts can also lead to avoided level crossings, we assume these to be typically off-resonant and thus irrelevant in most scenarios. In light of these two considerations, we do not include the effect of higher-order Zeeman effects. However, when considering the implementation of this protocol with a particular choice of states, it is important that these effects are taken into account.

Finally, we discuss the scaling of the phonon swap. In the case of the dipolar cooling, the scaling is the same as before: $\mathcal{G}_{\text{am}} \propto n^2/r^5$ or $\mathcal{G}_{\text{am}} \propto n^{-8}$ at half the LeRoy radius. 
The magnetic field primarily gives more freedom in state choice. However, the vdW cooling exhibits an improvement in scaling. In this case, the primary restriction is that the dipolar interactions remain smaller than the  atomic/molecular transition energy difference $C_3/r^3 \ll \delta_{\text{am}}$. This means that for any given separation, we can choose $\delta_{\text{am}} \propto C_3/r^3 \propto n^2/r^3$. As a result, $C_6/r^6 \propto (C_3/r^3)^2/\delta_{\text{am}} \propto C_3/r^3$, and the scaling behavior of the vdW interactions is that of the dipolar interactions, so $\mathcal{G}_{\text{am}}$ at the LeRoy radius falls off like $n^{-8}$ rather than $n^{-12}$.

\subsection{Interaction Form}

Now, we discuss how the form of the vdW interactions is changed due to the modified level structure, which shifts the relative intermediate state energies of the perturbative expansion. Depending on the strength of the magnetic field and the fine structure of the Rydberg states, the choice of quantum numbers can change from $J$ to $L$. To determine this, we can look at the Zeeman shift due to the magnetic field $\mu_B/\hbar B$ in the decoupled basis and compare this to fine structure as a function of $n$. In Table \ref{tab:fineZeeman}, we show the value of $n$ where they are equal at 10 mT for the P states. This illustrates that lithium and sodium have small fine structure that can more or less be neglected for any of the $n$ and magnetic fields under consideration. In comparison, for rubidium and cesium, the fine structure could be comparable to the Zeeman shifts, which can modify the form of the interactions. As illustrated in Fig.~\ref{fig:dressB}, $m_L = \pm 1$ can be used in the large magnetic field limit, both of which will give rise to the same form of interactions in the absence of fine structure. For a stretched state $m_L = 2 m_S$, the fine structure only enters as an energy shift, so we focus on this case for simplicity and expect the other option to exhibit similar behavior.

\begin{table}
\centering
\def\arraystretch{1}
\setlength\tabcolsep{2.4mm}
\begin{tabular}{ccccc}
\toprule
\toprule
 & $^6$Li & $^{23}$Na & $^{87}$Rb & $^{133}$Cs \\
\midrule
 $n$ & 8.2 & 34.5 & 88 & 119 \\
 \bottomrule
 \bottomrule
\end{tabular}
\caption{Value of $n$ at which fine structure is comparable to $\mu_B B/\hbar $ at $B = 10$ mT. \label{tab:fineZeeman}}
\end{table}

As usual, the dipolar interactions will exhibit the typical form of interaction. Here, we now focus on driving S$\to$P transitions for better comparison with vdW and since the tunability of the transition frequencies via the magnetic field reduces the advantage of the energy level asymmetries. To maximize the interaction strength, we consider driving $N, |m_N| = N$ to $N',|m_{N'}| = N'$ transitions, which requires $m_N, m_L$ have the same sign. For the vdW interactions, we again assume that only one manifold of states will dominate, particularly in light of the ability to tune the transition frequencies close via the magnetic field. We again focus on $|m_N| = N$ states. While it is possible to utilize $N= 1, m_N = 0$ while utilizing two drives to bring the $m_N = \pm 1$ states out of degeneracy, we find the strongest vdW interactions using $|m_N| = N$ states. Finally, we have a choice in the sign of $m_L$, which we take to be same as $m_N$ to maximize the vdW interactions.
For $N= m_N = 2$, we find the interaction form
\begin{subequations}
    \begin{equation}
        \mathcal{D}^{(1;3)} = \frac{253 - 264 \cos 2 \theta + 27 \cos 4 \theta}{5040}, 
    \end{equation}
    \begin{equation}
        \hat V_x \approx -\frac{34}{105}\frac{(\hat x_{\text{a}} - \hat x_{\text{m}})^2}{ r_{\text{am}}^5}, \qquad
        \hat V_y \approx -\frac{68}{105}\frac{\hat y_{\text{a}} - \hat y_{\text{m}}}{r_{\text{am}}^4} + \frac{34}{15} \frac{(\hat y_{\text{a}} - \hat y_{\text{m}})^2}{r_{\text{am}}^5}, \qquad 
        \hat V_z \approx -\frac{33}{70}\frac{(\hat z_{\text{a}} - \hat z_{\text{m}})^2}{r_{\text{am}}^5}.
    \end{equation}
\end{subequations}
For $N = m_N = 1$, we find
\begin{subequations}
    \begin{equation}
        \mathcal{D}^{(1;2)} = \frac{227 - 204 \cos 2 \theta + 9 \cos 4 \theta}{2880}, 
    \end{equation}
    \begin{equation}
        \hat V_x \approx -\frac{11}{24}\frac{(\hat x_{\text{a}} - \hat x_{\text{m}})^2}{ r_{\text{am}}^8}, \qquad
        \hat V_y \approx -\frac{11}{12}\frac{\hat y_{\text{a}} - \hat y_{\text{m}}}{r_{\text{am}}^7} + \frac{77}{24} \frac{(\hat y_{\text{a}} - \hat y_{\text{m}})^2}{r_{\text{am}}^8}, \qquad 
        \hat V_z \approx -\frac{5}{8}\frac{(\hat z_{\text{a}} - \hat z_{\text{m}})^2}{r_{\text{am}}^8}.
    \end{equation}
\end{subequations}
Finally, utilizing the $N=0$ states leads to
\begin{subequations}
    \begin{equation}
        \mathcal{D}^{(1;1)} = \frac{5 - 3 \cos^2 \theta }{18}, 
    \end{equation}
    \begin{equation}
        \hat V_x \approx -\frac{5}{6}\frac{(\hat x_{\text{a}} - \hat x_{\text{m}})^2}{ r_{\text{am}}^8}, \qquad
        \hat V_y \approx -\frac{5}{3}\frac{\hat y_{\text{a}} - \hat y_{\text{m}}}{r_{\text{am}}^7} + \frac{35}{6} \frac{(\hat y_{\text{a}} - \hat y_{\text{m}})^2}{r_{\text{am}}^8}, \qquad 
        \hat V_z \approx -\frac{(\hat z_{\text{a}} - \hat z_{\text{m}})^2}{r_{\text{am}}^8}.
    \end{equation}
\end{subequations}
Note that the reduced matrix elements in the $L$ basis should be used when calculating $C_6^{(p)}$ values in this case.

In Table \ref{tab:magnetic}, we present cooling parameters realizable with a magnetic field.
The $n$S column indicates the smallest value of $n$ for which a magnetic field of no more than 100 mT can make the Rydberg and molecular transition frequencies resonant with one another.
This choice of $n$ is used for all parameters up to (and including) $r_{0.95}$, from which values for higher $n$ can be obtained via usual scaling laws. Additionally, $r_{0.95}^{\text{max}}$ denotes the maximum value this can take before half the LeRoy radius becomes larger at a corresponding $n_{0.95}^{\text{max}}$S Rydberg state; level crossings and near-degeneracies may make the realization of these parameters unphysical. In light of this, we also report $n_B$, which defines the value of $n$ at which the necessary Zeeman shift is approximately the same as the atomic transition frequency, i.e., when the molecular transition frequency is twice that of the atomic transition frequency. This defined the cutoff in the main text. Finally, $\delta_{\text{am},0.05}$ denotes the transition frequency detuning that corresponds to $5\,\%$ of the perturbative pair-state being composed of intermediate states at half the LeRoy radius. We see that these are on the order of 10s to 100s of MHz. Utilizing the fact that these values scale with $C_3/r_{\text{am}}^{-3}$ and that $r_{0.95} \propto n$, then $\delta_{\text{am},0.05} \propto n^{-1}$. For the values considered here, we find that this transition frequency difference remains on the order of $2\pi \times 1$ MHz at its lowest, so precise fine-tuning of the magnetic field is unnecessary.

\begin{table}[h]
\centering
\setlength\tabcolsep{.95mm}
\begin{tabular}{ccccccccccccccc} \toprule \toprule
 & & $N$ & nS & $B_{\text{res}}$ & $n_B$ & $r^{-\alpha}$ & $|C_\alpha|/2 \pi$ &
 $\mathcal{G}_{\text{am},z}/2 \pi$ & $\mathcal{G}_{\text{am,z}}/\gamma_r$   & $a_{\text{LR}}/2$ & $r_{0.95}$ & $r_{0.95}^{\text{max}}$ & $n_{0.95}^{\text{max}}$ & $|\delta_{\text{am},0.05}|/2 \pi$ \\
 & & & & (mT) &  & (\SI{}{\kilo\Hz.\micro\metre^\alpha}) & (kHz) & [$a_{\text{LR}}/2$] & (\SI{}{\micro\metre}) & (\SI{}{\micro\metre}) &  
 (\SI{}{\micro\metre}) & (\SI{}{\micro\metre}) & & (MHz), [$a_{\text{LR}}/2$] \\
 \midrule
 & & \ \multirow{2}{*}[-0.5\dimexpr \aboverulesep + \belowrulesep + \cmidrulewidth]{1} \ 
 & \multirow{2}{*}[-0.5\dimexpr \aboverulesep + \belowrulesep + \cmidrulewidth]{76S} 
 & \multirow{2}{*}[-0.5\dimexpr \aboverulesep + \belowrulesep + \cmidrulewidth]{81.0}
 & \multirow{2}{*}[-0.5\dimexpr \aboverulesep + \belowrulesep + \cmidrulewidth]{100~}
 & 3 & $1003$ & 43.2 & 138 & 0.93 & 1.45 & 2.25 & 118 & \\
 \cmidrule{7-15}
 & \ \multirow{2}{*}[-0.5\dimexpr \aboverulesep + \belowrulesep + \cmidrulewidth]{Na} \ & & & & & 6 & $1101$ & 53 & 188 & 0.93 & 1.54 & 2.56 & 126 & $33$ \\
 \cmidrule{3-15}
 & & \ \multirow{2}{*}[-0.5\dimexpr \aboverulesep + \belowrulesep + \cmidrulewidth]{2} \ & \multirow{2}{*}[-0.5\dimexpr \aboverulesep + \belowrulesep + \cmidrulewidth]{68S} & \multirow{2}{*}[-0.5\dimexpr \aboverulesep + \belowrulesep + \cmidrulewidth]{67.6} & \multirow{2}{*}[-0.5\dimexpr \aboverulesep + \belowrulesep + \cmidrulewidth]{87} & 3 & $826$ & 111 & 252 & 0.74 & 1.30 & 2.28 & 119 &  \\
 \cmidrule{7-15}
 \multirow{2}{*}[-0.5\dimexpr \aboverulesep + \belowrulesep + \cmidrulewidth]{$^{23}$Na$^{133}$Cs} \ & &  & & & & 6 & $457$ & 144 & 364 & 0.74 & 1.40 & 2.64 & 128 & 54 \\
 \cmidrule{2-15}
 & & \ \multirow{2}{*}[-0.5\dimexpr \aboverulesep + \belowrulesep + \cmidrulewidth]{1} \ & \multirow{2}{*}[-0.5\dimexpr \aboverulesep + \belowrulesep + \cmidrulewidth]{79S} & \multirow{2}{*}[-0.5\dimexpr \aboverulesep + \belowrulesep + \cmidrulewidth]{77.5} & \multirow{2}{*}[-0.5\dimexpr \aboverulesep + \belowrulesep + \cmidrulewidth]{103} & 3 & $1007$ & 17.3 & 40.1 & 0.94 & 1.14 & 1.39 & 96 &   \\
 \cmidrule{7-15}
 & \ \multirow{2}{*}[-0.5\dimexpr \aboverulesep + \belowrulesep + \cmidrulewidth]{Cs} \ & &  & &  & 6 & $1132$ & 21.3 & 67.8 & 0.94 & 1.27 & 1.70 & 106 & $33$ \\
 \cmidrule{3-15}
 & & \ \multirow{2}{*}[-0.5\dimexpr \aboverulesep + \belowrulesep + \cmidrulewidth]{2} \ & \multirow{2}{*}[-0.5\dimexpr \aboverulesep + \belowrulesep + \cmidrulewidth]{70S} & \multirow{2}{*}[-0.5\dimexpr \aboverulesep + \belowrulesep + \cmidrulewidth]{99.0} & \multirow{2}{*}[-0.5\dimexpr \aboverulesep + \belowrulesep + \cmidrulewidth]{90} & 3 & $805$ & 49.8 & 78.6 & 0.73 & 1.01 & 1.40 & 97 & \\
 \cmidrule{7-15}
 & & &  & & & 6 & $418$ & 65.1 & 141 & 0.73 & 1.14 & 1.77 & 109 & 56 \\
\midrule
 & & \ \multirow{2}{*}[-0.5\dimexpr \aboverulesep + \belowrulesep + \cmidrulewidth]{1} \ 
 & \multirow{2}{*}[-0.5\dimexpr \aboverulesep + \belowrulesep + \cmidrulewidth]{55S} 
 & \multirow{2}{*}[-0.5\dimexpr \aboverulesep + \belowrulesep + \cmidrulewidth]{40.7}
 & \multirow{2}{*}[-0.5\dimexpr \aboverulesep + \belowrulesep + \cmidrulewidth]{69}
 & 3 & $481$ & 980 & 654 & 0.50 & 1.06 & 2.25 & 117 & \\
 \cmidrule{7-15}
 & \ \multirow{2}{*}[-0.5\dimexpr \aboverulesep + \belowrulesep + \cmidrulewidth]{Li} \ & &  & & & 6 & $81.0$ & 1207 & 1045 & 0.50 & 1.16 & 2.70 & 128 & $105$ \\
 \cmidrule{3-15}
 & & \ \multirow{2}{*}[-0.5\dimexpr \aboverulesep + \belowrulesep + \cmidrulewidth]{2} \ & \multirow{2}{*}[-0.5\dimexpr \aboverulesep + \belowrulesep + \cmidrulewidth]{48S} & \multirow{2}{*}[-0.5\dimexpr \aboverulesep + \belowrulesep + \cmidrulewidth]{86.2} & \multirow{2}{*}[-0.5\dimexpr \aboverulesep + \belowrulesep + \cmidrulewidth]{60} & 3 & $377$ & 2860 & 1267 & 0.38 & 0.92 & 2.22 & 116 & \\
 \cmidrule{7-15}
 \multirow{2}{*}[-0.5\dimexpr \aboverulesep + \belowrulesep + \cmidrulewidth]{$^{6}$Li$^{133}$Cs} \ & &  & & & & 6 & $27.7$ & 3739 & 2147 & 0.38 & 1.02 & 2.74 & 129 & 186 \\
 \cmidrule{2-15}
 & & \ \multirow{2}{*}[-0.5\dimexpr \aboverulesep + \belowrulesep + \cmidrulewidth]{1} \ & \multirow{2}{*}[-0.5\dimexpr \aboverulesep + \belowrulesep + \cmidrulewidth]{54S} & \multirow{2}{*}[-0.5\dimexpr \aboverulesep + \belowrulesep + \cmidrulewidth]{88.9} & \multirow{2}{*}[-0.5\dimexpr \aboverulesep + \belowrulesep + \cmidrulewidth]{67} & 3 & $501$ & 529 & 363 & 0.42 & 0.79 & 1.50 & 102 & \\
 \cmidrule{7-15}
 & \ \multirow{2}{*}[-0.5\dimexpr \aboverulesep + \belowrulesep + \cmidrulewidth]{Cs} \ & &  & & & 6 & $49.4$ & 651 & 612 & 0.42 & 0.88 & 1.83 & 113 & $186$ \\
 \cmidrule{3-15}
 & & \ \multirow{2}{*}[-0.5\dimexpr \aboverulesep + \belowrulesep + \cmidrulewidth]{2} \ & \multirow{2}{*}[-0.5\dimexpr \aboverulesep + \belowrulesep + \cmidrulewidth]{48S} & \multirow{2}{*}[-0.5\dimexpr \aboverulesep + \belowrulesep + \cmidrulewidth]{77.1} & \multirow{2}{*}[-0.5\dimexpr \aboverulesep + \belowrulesep + \cmidrulewidth]{59} & 3 & $400$ & 1434 & 672 & 0.32 & 0.68 & 1.46 & 102 & \\
 \cmidrule{7-15}
 & & &  & & & 6 & $18.2$ & 1874 & 1200 & 0.32 & 0.77 & 1.85 & 115 & 319 \\
 \midrule 
 & & \ \multirow{2}{*}[-0.5\dimexpr \aboverulesep + \belowrulesep + \cmidrulewidth]{0} \ & \multirow{2}{*}[-0.5\dimexpr \aboverulesep + \belowrulesep + \cmidrulewidth]{57S} & \multirow{2}{*}[-0.5\dimexpr \aboverulesep + \belowrulesep + \cmidrulewidth]{74.1} & \multirow{2}{*}[-0.5\dimexpr \aboverulesep + \belowrulesep + \cmidrulewidth]{72} & 3 & 306 & 288 & 237 & 0.49 & 0.84 & 1.46 & 99 & \\
 \cmidrule{7-15}
   \multirow{2}{*}[-0.5\dimexpr \aboverulesep + \belowrulesep + \cmidrulewidth]{$^{40}$Ca$^{19}$F} & \ \multirow{2}{*}[-0.5\dimexpr \aboverulesep + \belowrulesep + \cmidrulewidth]{$^{87}$Rb} \ & & & &  & 6 & 49.5 & 326 & 400 & 0.49 & 0.93 & 1.80 & 110 & 76 \\
 \cmidrule{3-15}
 & & \ \multirow{2}{*}[-0.5\dimexpr \aboverulesep + \belowrulesep + \cmidrulewidth]{1} \ & \multirow{2}{*}[-0.5\dimexpr \aboverulesep + \belowrulesep + \cmidrulewidth]{47S} & \multirow{2}{*}[-0.5\dimexpr \aboverulesep + \belowrulesep + \cmidrulewidth]{135.3} & \multirow{2}{*}[-0.5\dimexpr \aboverulesep + \belowrulesep + \cmidrulewidth]{57} & 3 & $221$ & 1623 & 723 & 0.32 & 0.70 & 1.51 & 102 & \\
 \cmidrule{7-15}
 & & &  & & & 6 & $10.0$ & 1998 & 1326 & 0.32 & 0.79 & 1.93 & 115 & $180$ \\
\bottomrule \bottomrule
\end{tabular}
\caption{Parameters for hyperfine cooling with a magnetic field. We assume $\omega_z = 2 \pi \times 25$ kHz. Parentheses and square brackets denote units and the value of $r_{\text{am}}$ where the parameter was calculated, respectively, when applicable. Absolute values are used since the sign of $C_3$ is determined by the choice of even/odd superposition and $C_6$ can take either sign depending on the sign of $\delta_{\text{am},0.05}$, which can likewise be tuned via the magnetic field. The summarized data in Fig.~4 of the main text correspond to linear plots (in $n$) between $r_{0.95}$ and $r_{0.95}^{\text{max}}$ truncated at the smaller of $n_B$, $n_{0.95}^{\text{max}}$. \label{tab:magnetic}}
\end{table}

\section{DC Cooling}
\label{sec:dc}
In this section, we discuss in further detail how a DC electric field could in principle be utilized for cooling. In this case, the electric field induces permanent dipole moments in the atomic and molecular states, realizing a direct interaction between the two. Here, we encode the molecular states in hyperfine states adiabatically connected to $N=0$. 

There are two primary regimes to consider for the strength of the DC field. The first is the regime where the electric field has a largely perturbative effect on the Rydberg states, so there are minimal level crossings and the auxiliary cooling state has high overlap with the bare states in the absence of an electric field. This limits the possible electric fields to $E \propto n^{-5}$ according to the Inglis-Teller limit, which is defined by the onset of level crossings of different $n$ states \cite{Gallagher1994}. Given the smaller dipole moment of the molecule, the electric field is always perturbative. 

Going beyond the Inglis-Teller limit, the various Rydberg levels begin to strongly mix in a complex Stark map. In this case, the next limitation on the electric field is described by the classical ionization limit: the electric field at which the electron is expected classically to ionize \cite{Gallagher1994}. This is defined as the electric field at which the saddle point of the full Coulomb and electric field potential occurs at the energy of the atomic state, indicating that the electron is no longer classically bound. Beyond this limit, the atomic states have ionization rates which rapidly increase with the electric field. The possible electric fields scale as $E \propto n^{-4}$, so the extent of state mixing increases at higher $n$. Note that in this limit, the Stark shifts are very strong, and the classical ionization limit scaling is better described as the electric field scaling with the square of the bound state energy.

In principle, it is possible to go beyond this limit by up to a factor of 2 in the electric field, and Rydberg blockade on the order of and beyond the classical ionization limit has been experimentally explored \cite{Stecker2020}. However, this relies on careful tuning of level crossings to realize destructive interference of the ionization rates. In light of this and the fact that we expect limited gains in $\mathcal{G}_{\text{am}}/\gamma_r$ at the higher fields, we do not expect this to be a viable approach for cooling and consider only fields weaker than the classical ionization limit.

\begin{table}[b]
\centering
\def\arraystretch{1}
\setlength\tabcolsep{3.1mm}
\begin{tabular}{cccccccccc}
\toprule
\toprule
  & & $n$ & $E_{\text{opt}}$ & $a_{\text{LR}}/2$  & $\mathcal{G}_{\text{am},z}/2 \pi$ & $\mathcal{G}_{\text{am},z}/\gamma_r$ & $\mathcal{G}_{\text{am},z}/2 \pi$ & $\mathcal{G}_{\text{am},z}/\gamma_r$ & $r_{0.95}$  \\
  & & & (V/cm) & (\SI{}{\micro\metre}) &  (kHz), [$a_{\text{LR}}/2$] & [$a_{\text{LR}}/2$] &  (kHz), [1 \SI{}{\micro\metre}] & [\SI{1}{\micro\metre}] & (\SI{}{\micro\metre}) \\
\midrule
  & & 20 & 2612 & 0.076 & 11.0 & $2.22$ & $4.23 \times 10^6$ & $8.53 \times 10^5$ & 0.684 \\
 & &  35 & 235 & 0.21 & 2.16 & 3.32 & $5.80 \times 10^3$ & $8.90 \times 10^3$ & 0.742 \\
 & Na &  50 & 57.2 & 0.42 & 0.830 & 3.48 & 62.9 & 264 & 0.749 \\
 & &  65 & 19.5 & 0.72 & 0.593 & 3.97 & 3.07 & 20.6 & 0.768 \\
\multirow{2}{*}[-0.5\dimexpr \aboverulesep + \belowrulesep + \cmidrulewidth]{$^{23}$Na$^{133}$Cs} \ & & 80 & 7.97 & 1.18 & 0.29 & 4.34 & 0.128 & 1.91 & 0.783 \\
 \cmidrule{2-10}
& & 20 & 4860 & 0.12 & 5.43 & 1.71 & $2.08 \times 10^5$ & $6.53 \times 10^4$ & 0.649 \\
 & & 35 & 347 & 0.17 & 3.16 & 1.73 & $2.20 \times 10^4$ & $1.20 \times 10^4$ & 0.651 \\
 & Cs & 50 & 53.7 & 0.35 & 1.18 & 2.15 & 209 & 382 & 0.680\\
 & & 65 & 14.9 & 0.58 & 0.352 & 2.48 & 5.56 & 39.1 & 0.700\\
 & & 80 & 6.95 & 0.91 & 0.210 & 2.31 & 0.342 & 3.76 & 0.690 \\
 \midrule
 & & 20 & 2148 & 0.096 & 6.43 & 2.29 & $7.88 \times 10^5$ & $2.80 \times 10^5$ & 0.689 \\
 & & 35 & 191 & 0.21 & 2.12 & 2.53 & 5430 & 6470 & 0.703\\
 & Li & 50 & 48.2 & 0.41 & 0.631 & 2.20 & 57.5 & 201 & 0.683 \\
 & & 65 & 17.6 & 0.79 & 0.331 & 2.09 & 1.05 & 6.61 & 0.676\\
 \multirow{2}{*}[-0.5\dimexpr \aboverulesep + \belowrulesep + \cmidrulewidth]{$^{6}$Li$^{133}$Cs} \ & & 80 & 7.92 & 1.13 & 0.226 & 2.51 & 0.124 & 1.38 & 0.702 \\
 \cmidrule{2-10}
 & & 20 & 4860 & 0.12 & 2.15 & 0.674 & $8.23 \times 10^4$ & $2.58 \times 10^4$ & 0.539 \\
 & & 35 & 347 & 0.17 & 1.25 & 0.684 & 8709 & 4762 & 0.541 \\
 & Cs & 50 & 53.8 & 0.35 & 0.465 & 0.848 & 82.7 & 151 & 0.565\\
 & & 65 & 14.9 & 0.58 & 0.139 & 0.980 & 2.20 & 15.5 & 0.581\\
 & & 80 & 6.95 & 0.91 & 0.0828 & 0.911 & 0.135 & 1.49 & 0.573\\
 \bottomrule
 \bottomrule
\end{tabular}
\caption{Optimized DC Cooling parameters of NaCs and LiCs for maximal $r_{0.95}$ values. Parentheses and square brackets denote units and the value of $r_{\text{am}}$ where the parameter was calculated, respectively, when applicable. \label{tab:dc}}
\end{table}

In order to examine the performance of dc cooling for a given choice of $n$, we adopt the following approach. First, we restrict the Hilbert space to $n- \delta n \leq n \leq n + \delta n$, for $\delta n$ up to 22 for the highest choice of $n$, and $L \leq 22$. Second, we diagonalize the resulting Hamiltonian for electric fields up to the classical ionization limit, which we define with respect to the energy of the $n$S state in the absence of an electric field. Third, we restrict the energy window of the potential states to those which are closer to the zero field $n$S state energy than that of the $(n-1)$S or $(n+1)$S states. 
Finally, we diagonalize the molecular Hamiltonian in the presence of an electric field to determine the dipole moment of the $N=0$ rotational state and identify the optimal electric field for the largest value of $r_{0.95}$. In Table \ref{tab:dc}, cooling parameters which realize optimal $r_{0.95}$ are presented. While there is a degree of improvement in $r_{0.95}$, it is minimal, indicating the ability to go beyond the Ising-Teller limit leads to relatively small improvements at high $n$. Correspondingly, the scaling of the optimal electric field strengths tends to be closer to $n^{-5}$ like the Ising-Teller limit rather than the $n^{-4}$ scaling of the classical ionization limit.

\section{Fidelity due to deviations from identity}
\label{sec:fidelity}
In this section, we consider how fidelity is modified by deviations from identity interactions. For simplicity, we take the interactions to be $(I + \epsilon Z) V(r,\theta)$ with a proper choice of basis that we call $|\tilde{0}\rangle, |\tilde{1}\rangle$. We take the initial state to be $ |\psi(0) \rangle = a |\tilde{0},\sqrt{n},0\rangle + b |\tilde{1},\sqrt{n},0\rangle$, where the latter two arguments of the states denote the phonons for the molecule and atom in a coherent state basis, where we assume $n$ initial molecular phonons, although other states (e.g., Fock states) may be more physically relevant depending on the heating processes.
After a time $t = \pi/2$ for just a molecule-atom pair, the phonon exchange Hamiltonian causes the state to evolve to
\begin{equation}
    |\psi(\pi/2)\rangle =  a \left|\tilde{0}, \epsilon \sqrt{n}, \sqrt{(1-\epsilon) n } \right\rangle + b \left|\tilde{1}, -\epsilon \sqrt{n}, \sqrt{(1-\epsilon) n } \right\rangle.
\end{equation}
Here, we can immediately trace out the Rydberg phonons since they are not entangled, although even if they were entangled (e.g., from a slightly different interaction time), the relative shift of the two coherent states will be proportional to $\epsilon$, and there will be no changes to the scaling we identify. 

Next, we trace out the molecular phonons to find the molecular density matrix 
\begin{equation}
    \rho_{\text{m}} = \left( \begin{array}{cc}
        |a|^2 & a b^* e^{-2 n \epsilon^2} \\
        a^* b e^{-2 n \epsilon^2} & |b|^2,
    \end{array}
    \right),
\end{equation}
whose fidelity with $a |\tilde{0}\rangle + b |\tilde{1}\rangle$ is given by
\begin{equation}
    1 - 4|a|^2 (1 - |a|^2) e^{- n \epsilon^2} \sinh \epsilon^2 \approx 1 - 4|a|^2 (1 - |a|^2) n \epsilon^2,
\end{equation}
indicating that the state infidelity is quadratic in $\epsilon$ and linear in the number of phonons. However, it is worth pointing out that if multiple cooling steps or continuous cooling are utilized instead, then we would need to introduce a more specific form of heating in the dynamics.

\bibliography{MoleculeRydberg}